\documentclass[aps,pra,amsmath,twocolumn,showpacs,bibnotes,10pt]{revtex4-1}
\usepackage{epsfig,graphicx}

\begin{document}

\title{Surface states of gapped electron systems and semi-metals}

\author{Xin-Zhong Yan$^{1,2}$ and C. S. Ting$^3$}
\address{$^1$Beijing National Laboratory for Condensed Matter Physics, Institute of Physics, Chinese Academy of Sciences, Beijing 100190, China}
\address{$^2$Beijing Key Laboratory for Advanced Functional Materials and Structure Research, Beijing 100190, China} 
\address{$^3$Texas Center for Superconductivity, University of Houston, Houston, Texas 77204, USA}

\date{\today}

\begin{abstract}
With a generic lattice model for electrons occupying a semi-infinite crystal with a hard surface, we study the eigenstates of the system with a bulk band gap (or the gap with nodal points). The exact solution to the wave functions of scattering states is obtained. From the scattering states, we derive the criterion for the existence of surface states. The wave functions and the energy of the surface states are then determined. We obtain a connection between the wave functions of the bulk states and the surface states. For electrons in a system with time-reversal symmetry, with this connection, we rigorously prove the correspondence between the change of Kramers degeneracy of the surface states and the bulk time-reversal $Z_2$ invariant. The theory is applicable to systems of (topological) insulators, superconductors, and semi-metals. Examples for solving the edge states of electrons with/without the spin-orbit interactions in graphene with a hard zigzag edge and that in a two-dimensional $d$-wave superconductor with a (1,1) edge are given in appendices.
\end{abstract}

\maketitle
%\ioptwocol

\section{Introduction}

Surface states (SSs) of electron systems \cite{Tamm,Shockley} can exist in crystals of superconductors \cite{Thuneberg,Hu,Sauls}, semi-metals \cite{Fujita,Nakada,Wakabayashi}, and topological insulators (TI) \cite{Murakami,Sinova,Kane,Qi,HZhang,Hsieh,Hasan}. The existence of SSs in interface between metals or superconductors leads to sizable electronic tunneling \cite{Tanaka,Yan,Wimmer}. The TI is characterized by the existence of the conducting SSs and an insulating bulk gap generated by the spin-orbit interactions (SOI). The materials existing SSs have prospective applications in electronic/spintronic devices. In particular, the TI can be used to conduct spins on the surface due to spin-Hall effect so that there is no electric resistance and no energy cost \cite{Murakami,Sinova,Kane,Kane2,Sheng,Wu,Xu,Bernevig2,Konig}.

The existence of SSs in TI is considered as topologically protected. There is a correspondence between the existence of SSs and the topological property of the bulk states. For classification of the TI on the basis of the topological invariant \cite{Thouless1,Thouless}, Kane {\it et. al.} \cite{Kane2,Fu,Fu2,Fu3} introduce the time reversal (TR) polarization and define the $Z_2$ topological invariant for the bulk states. The $Z_2$ topological invariant is interpreted using a Laughlin-type fictitious experiment \cite{Laughlin}. In the Laughlin's construction for the quantum Hall effect on a cylinder, the change of magnetic flux inserted in the cylinder can transfer electrons from one end to another through the cylinder. The charge transfer can be interpreted as a change in the charge polarization of the cylinder. Since the charge transfer corresponds to the change of the number of edge states \cite{Halperin,Hatsugai}, the charge polarization is a description of the number of edge states. In a TI, there are two type states for electrons because of the Kramers degeneracy. In analogous to the quantum Hall effect on a cylinder, the TR polarization is associated with the Kramers degeneracy of the two type edge states. On the other hand, the change of the momentum parallel to the edge of the cylinder plays the role of the change of flux. Therefore, the $Z_2$ invariant being the change of the TR polarization between two invariant momenta parallel to the edge is interpreted as the change of the Kramers degeneracy of SSs between the two invariant momenta. For a three dimensional system, the $Z_2$ invariant is interpreted using the similar analogy with quantum Hall effect.  

The SSs and the bulk-boundary correspondence have been studied by many theoretical works using various concrete models including lattice \cite{Qi2,Hosur,Mong,Isaev,Delplace,Ryu,Pershoguba,Siroki,Zhao} and continuous models \cite{Linder,Liu,Zhang,Enaldiev,Weber}. In some of the existing works, the SSs are analytically solved by imposing special boundary conditions \cite{Isaev,Pershoguba}. The zero-energy edge states are studied for the Dirac fermions and topological superconductors \cite{Mong,Ryu}. For the continuous model, a set of the parameters for the boundary conditions needs to be determined by the bulk system and the crystal potential near the surface \cite{Enaldiev}. It is shown that the boundary conditions strongly affect the spectrum of the SSs and even the existence of the states near the zero momentum parallel to the surface. 

Most works study the bulk-boundary correspondence using analogy with the charge transfer by flux change in quantum Hall effect \cite{Kane2,Fu,Fu2,Fu3,Qi2,Roy,Roy2}. In fact, the $Z_2$ invariant is a characteristic of the bulk system rather than depending on the conditions of the crystal surface. After all, the edge states in the quantum Hall effect are the Landau edge states. They are certainly different from the SSs of a TI in the absence of the magnetic field. A question still remains as how to prove the bulk-boundary correspondence on a generic model with the TR symmetry without using analogy with the charge transfer by flux change. 

In this paper, with a generic lattice model, we present exact solution to the scattering states of an electron system occupying a semi-infinite crystal with a hard surface. With the result, we derive the criterion for the existence of the SSs and determine the eigenvalues and wave functions of the SSs. We get the connection between the wave functions of the bulk states and the SSs. For the electron systems with the TRS in TIs, with this connection, we prove the correspondence between the Kramers degeneracy of the SSs below the Fermi energy and the TR polarization introduced by Kane {\it et. al.} without using analogy with the charge transfer by flux change in quantum Hall effect. We present examples for solving the edge states of electrons with/without the SOI in graphene for a number of cases and that in a two-dimensional $d$-wave superconductor.

For concisely illustrating the physics, we present the theory for the simple band structure in the main text. We extend the analysis to the cases of more complicated band structures in the appendices. 

\section{Bulk States} 

We consider an electron system occupying a semi-infinite lattice with a hard surface. The $x$ axis of the coordinates is set as along the inner normal direction of the surface. The unit cell of the lattice contains $n$ atoms (or $n$ orbitals for an electron). The Hamiltonian of the system is given by
\begin{equation}
H= \sum_{ij}C^{\dagger}_{i}H_{ij}C_{j}\nonumber\\
\end{equation}
where $C^{\dagger}_{i} = (c^{\dagger}_{i1\uparrow},c^{\dagger}_{i2\uparrow},\cdots,c^{\dagger}_{in\uparrow},c^{\dagger}_{i1\downarrow},c^{\dagger}_{i2\downarrow},\cdots,c^{\dagger}_{in\downarrow})$ with $c^{\dagger}_{ils}$ creating an electron of spin $s$ on $l$th atom of $i$th unit cell, $H_{ij}$ is a $2n\times 2n$ matrix, and the sum runs over the unit cells in the semi-infinite space $i_x \ge 1$ and $j_x \ge 1$. Since the momentum parallel to the surface, $k_{\parallel}$, is a good quantum number, we will work in the space of $k_{\parallel}$ but real space along the $x$ axis. In this space, the Hamiltonian $H_{ij}$ then reads $H(l,k_{\parallel})$ with $l = j_x-i_x$. We suppose that the electron hopping is confined within a range: $-L \le l \le L$. For brevity of description, we hereafter may suppress the argument $k_{\parallel}$ unless it leads to confusion it will be explicitly written out again. We will use the units in which $e = \hbar = 1$.

First, we study the bulk eigenstates of electrons in the infinite lattice. The bulk states can be used as the basis for investigating the eigenstates of the system with a surface. For the bulk states, we work in the momentum $k$ space. The wave function $y(k)$ and the energy $E(k)$ of the bulk states are determined by
\begin{equation}
H(k)y_{\alpha}(k) = E_{\alpha}(k)y_{\alpha}(k), ~~{\rm for}~ \alpha= 1,\cdots,2n             \label{s1}
\end{equation}
where $\alpha = 1,\cdots,2n$ means there are $2n$ energy bands. The transpose of the wave function $y(k)$ is expressed as
\begin{eqnarray}
y^{t}(k) &=& (u_1,u_2,\cdots,u_n,v_1,v_2,\cdots,v_n)/N_k  \nonumber\\
&\equiv& w^t(k)/N_k \label{wv}
\end{eqnarray}
where $u$'s and $v$'s are the components and $N_k$ is the normalization constant given by
\begin{equation}
N_k = (\sum_{i=1}^n|u_i|^2+\sum_{i=1}^n|v_i|^2)^{1/2}. \nonumber
\end{equation}

Here, we analyse the property of the wave function. Define the variable $z = \exp(ik_x)$
with $k_x$ as the momentum along $x$ axis. Since $H(k)$ is the Fourier transform of $H(-l,k_{\parallel})$ with $-L \le l \le L$, it can be written as $H(k) = z^{-L}h(z)$ where $h(z)$ is a $2n\times 2n$ matrix; each element of $h(z)$ is a polynomial of $z$. The highest order of the polynomial among these elements is $2L$. The energy $E$ is determined by
\begin{equation}
\det [H(k) - E] = 0,             \nonumber
\end{equation}
which is equivalent to 
\begin{equation}
\det [h(z) - z^{L}E] = 0. \label{seq2}
\end{equation}
The left hand side is a polynomial of $z$ and $z^LE$ with their highest orders as respectively $4nL$ and $2n$. For a given momentum $k$ (and thereby $z$) there are $2n$ solutions to $E$. Each solution $E$ is a function of $z$. Denote the cofactor of the $\lambda$th element in the first row of matrix $h(z)-z^LE$ as $\Delta_{1\lambda}$, which is a polynomial of $z$ and $z^LE$. A solution to the components $u$'s and $v$'s can be obtained as 
\begin{eqnarray}
u_{\lambda} = \Delta_{1\lambda}, ~
v_{\lambda} = \Delta_{1\lambda+n}, ~~ {\rm for}~ 1 \le \lambda \le n. \nonumber
\end{eqnarray}
This solution to equation (\ref{s1}) can be checked with equation (\ref{seq2}). So far, these results are defined on the unit circle $|z| = 1$ in the $z$ plane. 

We now take analytical continuation of the results in the $z$ plane. Note that there are $2n$ branches of $E$ for $z$ varying in the $z$ plane. For the one to one correspondence between $E$ and $z$, we need to consider one $z$ plane for each energy band; the $2n$ energy bands correspond to $2n$ $z$ planes. Clearly, $E$ is an analytical function of $z$ except at $z=0$ because of $h(z)$ being analytical in the given $z$ plane. In the limit $z \to 0$, $z^{L}E$ should be finite or vanish to satisfy equation (\ref{seq2}). Therefore, we have $E \propto z^{-L}$ at most in the limit $z \to 0$. Thus, the cofactors $\Delta_{1\lambda}$'s and thereby the wave function $w_k$ defined in equation (\ref{wv}) are analytical functions of $z$ in the $z$ plane.  Note that the components $u$'s and $v$'s can be multiplied with an analytical function of $z$. The resulted wave function still satisfies equation (\ref{s1}).

On the other hand, in general, the normalization constant $N_k$ is not an analytical function of $z$. In $N_k$,  for example, $|u_{\lambda}|^2 = u_{\lambda}u_{\lambda}^{\ast}$, the derivative of $u_{\lambda}^{\ast}$ with respect to $z$ does not satisfies the Cauchy-Riemann equation.

\section{Eigenstates of Electrons in Semi-Infinite Space}

Studying the eigenstates in the semi-infinite space is essentially a one-dimensional problem. In this section, we proceed the analysis starting from the scattering states of the electrons in a given band without any degeneracy.   

{\it Incoming and outgoing waves.} An incoming wave to the surface with wavenumber $k$ is defined as the wave with negative velocity $v_k = dE/dk < 0$. Accordingly, an outgoing wave is defined as that of positive velocity. The evanescent waves are those waves with their amplitudes damping as the distance from the surface increasing. 

Since the energy is a periodic continuous function of the momentum, we suppose there are $m$ maxima (and also $m$ minima) in an energy band $E(k)$ within the period $-\pi < k \le\pi$ and all the maxima (minima) have the same value $E_M~ (E_m)$. Therefore, we have $E_m \le E(k) \le E_M$. Such an energy band is shown in the top panel of figure 1. The number of the incoming waves is the same as the outgoing waves. The total number of the incoming and outgoing waves at a given energy $E$ is $2m$, which is the number of the independent plane waves of bulk states degenerated at the same energy $E$. Under the $z$ mapping, the $2m$ incoming and outgoing wavenumbers at energy $E$ are mapped to $2m$ points on the unit circle $|z| = 1$ in the $z$ plane.   

\begin{figure}[t]
\centerline{\epsfig{file=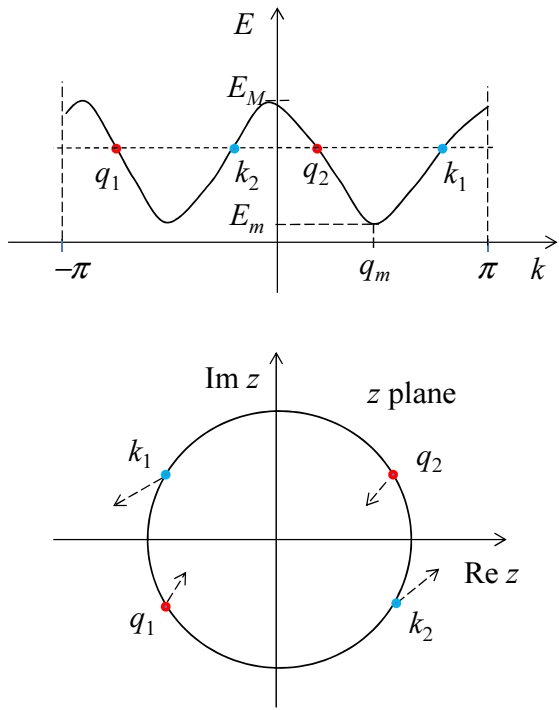,width=8.cm}}
\caption{(color online) Top panel: Sketch of an energy band. The wavenumbers of incoming and outgoing waves are denoted as red and blue dots, respectively. Lower panel: Under the mapping $z = \exp(iq)$, the points on the energy curve in the top panel are mapped onto the unit circle. Under analytical continuation with the points of incoming waves move to inside the unit circle, the points of outgoing waves go outside the unit circle.} 
\end{figure}

For later use, here, we consider the property of the incoming and outgoing waves under the analytical continuation with changing their real wavenumbers to complex wavenumbers. Consider the energy $E$ to be analytically continued to complex energy $E \to E + i\epsilon$ with $\epsilon$ as a infinitesimal small quantity. Then, the momentum $k$ is changed to $k+i\delta$ with $\delta$ determined by
\begin{eqnarray}
E+i\epsilon = E(k+i\delta) = E(k)+i\frac{dE}{dk}\delta. \label{eng}
\end{eqnarray}
For a given $\epsilon$, because of the dependence of velocity $dE/dk$, the signs of $\delta$ for incoming and outgoing waves are opposite. Therefore, under the analytical continuation in the $z$ plane, with the points of incoming waves moving to inside (outside) the unit circle, the points of outgoing waves go outside (inside) it.

For briefly expressing the wave function of the scattering state, we hereafter denote the lattice coordinate $j_x$ simply as $j$, and $k_x$ as $k$. In real space, the bulk eigenstate of momentum $k$ is a plane wave with the wave function given by
\begin{eqnarray}
\phi_k(j) =  y(k)z_k^j,  \label{plw}
\end{eqnarray}
with $z_k = \exp(ik)$. For an evanescent wave of complex wavenumber $\tilde k = k_r + ik_i$ with $k_i > 0$, its wave function is denoted as 
\begin{eqnarray}
\varphi_{\tilde k}(j) = y({\tilde k})z_{\tilde k}^j.  \label{evw}
\end{eqnarray}

\subsection{Scattering states} 

The wave function of a scattering state in which the electrons freely move in the system is superposed by the incoming and outgoing and evanescent waves. Consider a wave with wavenumber $q_{\mu}$ is incoming to the surface. It can be reflected by the surface to all the degenerated outgoing waves of wavenumbers $k_{\nu}$'s. Besides these waves, there are evanescent waves by the reflection. To see this, we consider the solutions to equation (\ref{seq2}) for a given energy $E$. Note that the order of the polynomial in equation (\ref{seq2}) is $2L\times 2n = 4nL$. Besides the $2m$ waves with real wavenumbers in the band under consideration, there are $2(2nL-m)$ additional waves of complex wavenumbers. These waves can be considered as the analytical continuation of incoming and outgoing waves of other bands. The $2nL-m$ (half total) complex wavenumbers have positive imaginary parts. These waves are the evanescent waves. All the rest $2nL-m$ waves of wavenumbers with negative imaginary parts are growing waves, not satisfying the boundary condition at $j \to \infty$. These growing waves should be ignored in the scattering waves. 

{\it Wave functions of scattering states.} With the above discussion and the definitions, we are ready to express the wave function of the scattering state of the incoming wave with wavenumber $q_{\mu}$ reflected to all the outgoing waves $k_{\nu}$'s and the evanescent waves $\tilde k_{\lambda}$'s all degenerated at energy $E$. For brevity, we will denote $\phi_{q_{\mu}}(j),~\phi_{k_{\nu}}(j)$ and $\varphi_{\tilde k_{\lambda}}(j)$ simply as $\phi_{\mu}(j),~\phi_{\nu}(j)$ and $\varphi_{\lambda}(j)$, respectively. Up to a normalization constant, the wave function of this state is given by 
\begin{eqnarray}
\psi_{\mu}(j;E) &=& \phi_{\mu}(j)-\sum_{\nu=1}^m\phi_{\nu}(j)S_{\nu\mu}-\sum_{\lambda=1}^{2nL-m}\varphi_{\lambda}(j)D_{\lambda\mu},\nonumber\\
  \label{swv}\\
 &&~~~~{\rm for~} \mu = 1,2,\cdots,m,~{\rm and}~j \geq 1, \nonumber\\
 &&~~~~{\rm with~} dE/dq_{\mu}\cdot dE/dk_{\nu} <0 \nonumber
\end{eqnarray}
where $S_{\nu\mu}$ and $D_{\lambda\mu}$ are constants depending on the energy $E$. 

The above analysis is also applicable to the process that a single outgoing wave comes from all the incoming waves and the evanescent waves under the surface reflections. The wave function is given by the similar formula as equation (\ref{swv}) where the wavenumbers $q_{\mu}$ and $k_{\nu}$'s are understood as the outgoing (with $dE/dq_{\mu} > 0$) and incoming ($dE/dk_{\nu} < 0$) waves, respectively. We can then allow the wavenumber $q$ varies in the whole range $(-\pi,\pi)$.
 
For the sake of description, we collect the wave functions in matrix forms as
\begin{eqnarray}
\psi(j,E) &=& [\psi_1(j,E),\psi_2(j,E),\cdots,\psi_m(j,E)] \nonumber\\
\phi_{\vec k}(j) &=& [\phi_1(j),\phi_2(j),\cdots,\phi_m(j)] \nonumber\\
\varphi_{\vec{\tilde k}}(j) &=& [\varphi_1(j),\varphi_2(j),\cdots,\varphi_{2nL-m}(j)]\nonumber 
\end{eqnarray}
where the vector $\vec k$ stands for $(k_1,\cdots,k_m)$ and the vector $\vec{\tilde k}$ for $(\tilde k_1,\tilde k_2,\cdots,\tilde k_{2nL-m})$ . Now, the formula (\ref{swv}) can be written in the compact form
\begin{eqnarray}
\psi(j;E) &=& \phi_{\vec q}(j)-\phi_{\vec k}(j)S-\varphi_{\vec{\tilde k}}(j)D,   
\label{cwv}
\end{eqnarray}
where $S$ is a matrix of dimension $m\times m$ and $D$ is a matrix of dimension $(2nL-m)\times m$. 

Because all the incoming and outgoing and evanescent waves in the wave function given by equation (\ref{cwv}) satisfy the Schr\"odinger equation in the bulk region ($j > L$), the wave function $\psi(j,E)$ satisfies the equation too,
\begin{eqnarray}
\sum_{l=-L}^{L}H(l)\psi(j+l;E) &=& E\psi(j;E),  ~~{\rm for}~j>L. \label{sdg0} 
\end{eqnarray} 
The matrices $S$ and $D$ should be determined by the boundary condition at the surface. 

{\it Boundary condition.} 
The surface of the lattice is defined as the truncation of the hopping. The electron cannot go out the lattice because of the vanishing of the hopping to a site outside the lattice. Close to the surface as shown in figure 2, for $1<j \le L$, the Schr\"odinger equation is different from equation (\ref{sdg0}). Now, it is given by 
\begin{eqnarray}
\sum_{l=1-j}^{L}H(l)\psi(j+l;E) &=& E\psi(j;E),   \label{sdg}\\
&&~~{\rm for}~j=1,2,\cdots,L.\nonumber
\end{eqnarray}
This is the boundary condition to the wave function at the surface. For the convenience of using this boundary condition, we rewrite equation (\ref{sdg}). By extending the definition of wave function given by equation (\ref{swv}) to $1-L \le j \le 0$, we write the left hand side of equation (\ref{sdg}) as
\begin{eqnarray}
\sum_{l=1-j}^{L}H(l)\psi(j+l;E)&&=\sum_{l=-L}^{L}H(l)\psi(j+l;E)\nonumber\\
&&-\sum_{l=-L}^{-j}H(l)\psi(j+l;E).    \label{sdg1}
\end{eqnarray}
Note that the first term in the right hand side of equation (\ref{sdg1}) equals $E\psi(j;E)$. From equations (\ref{sdg}) and (\ref{sdg1}), we get
\begin{eqnarray}
\sum_{l=-L}^{-j}H(l)\psi(j+l;E) &=& 0,   \label{bc}\\
&&~~{\rm for}~j=1,2,\cdots,L.\nonumber
\end{eqnarray}
This is the equivalent form of equation (\ref{sdg}). We emphasize that from equation (\ref{bc}) one cannot assert the vanishing of the wave function at sites $1-L \le j \le 0$. 

\begin{figure}[t]
\centerline{\epsfig{file=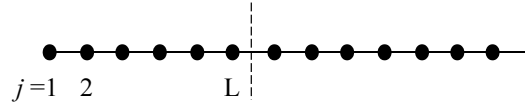,width=7.5cm}}
\caption{One-dimensional lattice with a left end. The black dots denote the lattice sites. For $1 \le j \le L$, the Schr\"odinger equation is different from that in the bulk region because there is no hopping between site-$j$ and a site outside the end.} 
\end{figure}

{\it Remark.} The surface of the lattice system is defined as the truncation of the hopping rather than an infinitive potential barrier. In real space, the wave function of an electron is defined within the lattice $j \ge 1$. Equation (\ref{bc}) says how the wave function would be if an electron went out the surface. 

To proceed, we define the matrix 
\begin{eqnarray}
h_j(k)&=&\sum_{l=-L}^{-j}H(l)\exp[ik(l+j)],   \label{bc1}\\
&&~~{\rm for}~j=1,2,\cdots,L\nonumber
\end{eqnarray}
and matrices
\begin{eqnarray}
h(\vec k) &=& \begin{bmatrix}
	h_1(k_1)&\cdots&h_1(k_m)\\
	\cdots&\cdots&\cdots\\
	h_L(k_1)&\cdots&h_L(k_m)\\
\end{bmatrix},           \label{mth} \\
Y(\vec k) &=& \begin{bmatrix}
	y(k_1)&0&\cdots\\
	0&y(k_2)&\cdots\\
	\cdots&\cdots&\cdots\\
	0&\cdots&y(k_m)\\
\end{bmatrix},             \label{mtw}\\
A(\vec k)&=& h(\vec k)Y(\vec k). \label{mta}
\end{eqnarray}
The dimensions of $h(\vec k),~Y(\vec k)$ and $A(\vec k)$ are $2nL\times 2nm,~2nm\times m$, and $2nL\times m$, respectively. Similarly, we can define the corresponding matrices for the evanescent waves. With these definitions, the $L$ equations (\ref{bc}) for $j = 1,2,\cdots,L$ can be written in a compact form
\begin{eqnarray}
A(\vec q)-A(\vec k)S-A(\vec{\tilde k})D = 0. \label{mtbc}
\end{eqnarray}
By denoting
\begin{eqnarray}
A_c &=& [A(\vec k),A(\vec{\tilde k})]. \nonumber\\
X &=&\begin{pmatrix}
	S \\
	D \\
\end{pmatrix},                         \nonumber  
\end{eqnarray}
equation (\ref{mtbc}) reads,
\begin{eqnarray}
A_cX=A(\vec q). \label{bcc}
\end{eqnarray}
Since the $m$ wave functions of real wavenumbers as the diagonal elements in $Y({\vec k})$ and the $2nL-m$ evanescent wave functions in $Y(\vec{\tilde k})$ are all independent, the $2nL$ columns of matrix $A_c$ are linearly independent vectors. Therefore, the matrix $A_c$ is invertible. We thus have 
\begin{eqnarray}
X = A_c^{-1}A(\vec q), \label{slt}
\end{eqnarray}
which is the solution to the constants $S$ and $D$. 

Here, we go further to analyse the solution. Since the diagonal elements (column vectors) of $Y(\vec q)$ are $m$ independent wave functions, the $m$ columns of $A(\vec q)$ are thereby independent vectors. Therefore, the rank of $A(\vec q)$ is $m$. According to the algebra theory, there exist $m$ independent row vectors in $A(\vec q)$. The other $2nL-m$ row vectors depend linearly on these $m$ independent vectors. These $2nL-m$ vectors can be eliminated by a row transformation $R(\vec q)$. Therefore, $A(\vec q)$ can be transformed to matrix $\underline{M}(\vec q)$ with $\underline{M}(\vec q)$ as a square block matrix $M(\vec q)$ of dimension $m\times m$ of rank $m$ and the rest block as zero matrix, 
\begin{eqnarray}
R(\vec q)A(\vec q)=\underline{M}(\vec q)\equiv\begin{bmatrix}
	M(\vec q)\\
	0\\
\end{bmatrix}.  
\label{sltb}
\end{eqnarray}
Thus, the solution to $X$ can be rewritten as
\begin{eqnarray}
X = A_c^{-1}R^{-1}(\vec q)\underline{M}(\vec q). \label{sltx}
\end{eqnarray}
Equation (\ref{sltx}) is the {\it central result} of this paper.

{\it Remark.} Recall $y(k)=w(k)/N_k$ defined by equation (\ref{wv}). The matrix $Y(\vec k)$ can be written as $Y(\vec k) = W(\vec k)f(\vec k)$ with $W(\vec k)$ defined similarly as by equation (\ref{mtw}) with the components $y(k)$'s replaced by $w(k)$'s and $f(\vec k) = diag(1/N_{k_1},1/N_{k_2},\cdots,1/N_{k_m})$. The diagonal matrix $f(\vec k)$ [$f(\vec{\tilde k})$] can be absorbed to the unknown matrix $S$ [$D$] in equation (\ref{mtbc}) by left product. The diagonal matrix $f(\vec q)$ from $A(\vec q)$ can also be absorbed to $S$ and $D$ by right product $f^{-1}(\vec q)$ to both sides of equation (\ref{mtbc}). All these mean that we can consider the components $y$'s in the matrix $Y$ simply as $w$'s. As indicated in sectin 2, the factor $1/N_k$ is not the analytical function of $z = \exp(ik)$. By absorbing these factors to the unknown coefficients, we get $A(\vec k) = h(\vec k)W(\vec k)$. The advantage of using $w$ is due to its simple analytical property. Hereafter, we will use this simplification.  

The $\mu$th column of $A(\vec q)$ is a function of $q_{\mu}$. Since all the momenta $q_{\mu}$'s are related with the same energy $E$, $E(q_{\mu}) = E(q_1)$, we can consider $A(\vec q)$ as a function of $q_1$. Using the variable $z \equiv \exp(iq_1)$, $A(\vec q)$ can be considered as a function of $z$ in the $z$ plane. Notice that
\begin{eqnarray}
\frac{z}{z_{\mu}}\frac{dz_{\mu}}{dz}= \frac{E(q_1)}{dq_1}/\frac{E(q_{\mu})}{dq_{\mu}}. \nonumber
\end{eqnarray}
The derivative $dz_{\mu}/dz$ exists only if the ratio in the right hand side of this equation exists. Though the $\mu$th column of $A(\vec q)$ is an analytical function of $z_{\mu}$ under the analytical continuation in the $z$ plane, as a function of $z$, it may be singular at some points where the derivative $dz_{\mu}/dz$ does not exist. Except for these singular points, $A(\vec q)$ can be analytically defined in the $z$ plane. Under the row transformation $R(\vec q)$, the obtained matrix $M(\vec q)$ has the same analytical property as $A(\vec q)$ in the $z$ plane. 

\subsection{Surface States} 

Now, we study the SSs. We note that the SSs are analogous to the bound states in a central force. Here, we analyse the SSs in the present problem with the similar approach for searching the bound states in the central force \cite{Levinson,Newton}.

We consider the case of changing the energy beyond the extrema of the band. We present the analysis for lowering the energy to below the lower bound $E_m$ of the band. The case of raising the energy above the upper bound can be analysed accordingly. As shown in Appendix A, the wavenumbers $q$'s and $k$'s become pairs of complex conjugates for $E < E_m$. For example, the wavenumbers $k_1$ and $q_2$ as shown in figure 1 move toward to $q_m$ in the process of lowering the energy in the band. Below the minimum $E_m$, they are $k_1 = q_m-iq_i$ and $q_2 = q_m+iq_i$. For $q_i > 0$, the wave of $q_2=q_m+iq_i$ is an evanescent wave with its amplitude damping as increasing the distance from the surface, while the wave of $k_1=q_m-iq_i$ is growing with distance. 

For $E < E_m$ and $q_i > 0$, the components of $\phi_{\vec q}(j)$ in the wave function $\psi(j;E)$ given by equation (\ref{cwv}) are evanescent waves. On the other hand, the components of $\phi_{\vec k}(j)$ in $\psi(j;E)$ are the growing waves as $j$ increasing. The wave function $\psi(j;E)$ does not satisfy the boundary condition at $j \to \infty$.
However, at certain energy, the determinant of $M(\vec q)$ may vanish 
\begin{eqnarray}
\det M(\vec q) = 0. \label{zrd}
\end{eqnarray}
Then, there exist a nonzero vector $a$ (of $m$ components) that  
\begin{eqnarray}
M(\vec q)a = 0. \label{zr}
\end{eqnarray}
From equation (\ref{sltx}), we then have $Xa =0$ and thereby $Sa = Da = 0$. Applying equation (\ref{zr}) to equation (\ref{cwv}), we get a wave function 
\begin{eqnarray}
\psi_b(j;E) = \phi_{\vec q}(j)a. \label{bs}
\end{eqnarray}
Therefore, the vector $a$ eliminates the growing waves and the evanescent waves of other bands. The obtained wave function given by equation (\ref{bs}) satisfies the boundary conditions at the surface and at $j \to \infty$.  Thus, the wave function $\psi_b(j;E)$ describes a SS. 

Equation (\ref{zrd}) is {\it the criterion for the existence of the SSs}. This is another central result of this work. In Appendix B, we show equation (\ref{zrd}) is the necessary and sufficient condition for the existence of the SSs.

\begin{figure}[t]
\centerline{\epsfig{file=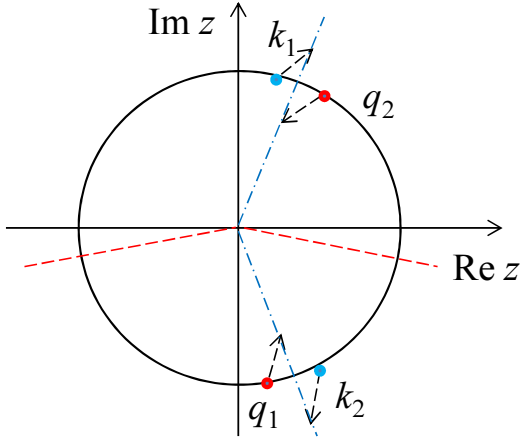,width=7.5cm}}
\caption{$z$ plane for the mapping $z = \exp(iq)$. For energy below (above) the lower (upper) bound of the band, the complex wavenumbers are mapped to the blue dash-dot (red dashed) lines. The angle of a blue dash-dot (red dashed) line is the momentum at the corresponding minimum (maximum) of the energy band.} 
\end{figure}

From equation (\ref{bs}), we get a conclusion that the SS consists of evanescent waves only belonging to the band under consideration. There is no any other evanescent wave from other bands in the components of the SS. This is different from the scattering states in which the evanescent waves from other bands are mixed in. This is an interesting result. 

In other words, the above conclusion can be expressed as that all the SSs of the system are classified by the continuous bands.

The matrix $M(\vec q)$ here takes the role of the generalised Jost function (matrix) in the theory of scattering states of electrons in a central force \cite{Levinson,Newton}. In that case, the existence of bound states is determined by the zeros of the determinant of the generalised Jost function. All the bound states are classified with the angular momentum. However, the situations are different in the two problems. Here, besides the scatterings between the incoming and the outgoing waves, there are scatterings from the incoming (outgoing) waves to the evanescent waves of other bands. On the other hand, in the central force problem, there are only the scatterings between the incoming and outgoing waves.

\subsection{Zeros of $\det M(\vec q)$}

Figure 3 shows the mapping $z = \exp(iq)$. With lowering the energy $E$ below the lower bound of the band, $z_{q_2}$ moves on the blue dash-dot line inside the unit circle. The angle of this blue dash-dot line equals $q_m$. On the other hand, $z_{k_1}$ moves on this line outside the unit circle. Since there are $m$ minima in the energy band, we get $m$ blue dash-dot lines in the $z$ plane. All $m$ pairs of $q$ and $k$ are then mapped to these $m$ blue dash-dot lines, respectively. For the energy above the upper bound of the band, the wavenumbers are mapped to the red dashed lines. The angle of each red dashed line is given by the momentum at the corresponding maximum of the energy band.

{\it Distribution of the zeros of $\det M(\vec q)$.} From the above analysis, we get that the zeros of $\det M(\vec q)$ are distributed on the blue dash-dot (red dashed) lines inside the unit circle for energy below (above) the lower (upper) bound of the band.

To count the number of the SSs of a band, we need to find out all the zeros of $\det M(\vec q)$. When $z = \exp(iq_1)$ arrives at a zero on its blue dash-dot line inside the unit circle, the other $m-1$ variables $z_{\mu}=\exp(iq_{\mu})~(\mu = 2,\cdots,m)$ respectively arrive at the zeros on their lines. Therefore, all these $m$ zeros correspond to a single SS. To count the number of zeros of $\det M(\vec q)$, one usually performs the contour integral in the complex $z$ plane. 

\begin{figure}[t]
\centerline{\epsfig{file=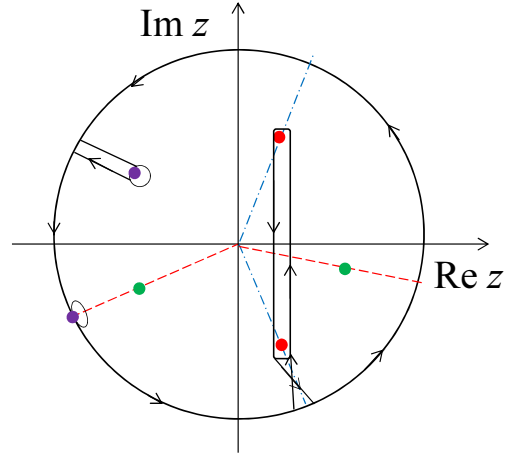,width=7.5cm}}
\caption{Contour $c$ in the $z$ plane for the integral in equation (\ref{ns}). The green (red) points are the possible regular (irregular) zeros of $\det M(\vec q)$. The purple points are the possible singular points of $\det M(\vec q)$, which are excluded from the interior of $c$.} 
\end{figure}

As noted above, the matrix $A(\vec q)$ and thereby the matrix $M(\vec q)$ are not simple analytical functions of $z$ in the unit circle. There my be singular points, which should be excluded from the interior of the contour. In addition, some of the zeros of $\det M(\vec q)$ inside the unit circle may be branch points. Close to these zeros $z_i~(i=1,2\cdots)$ of the branch points, $\det M(\vec q)$ behaves as $\det M(\vec q) \propto (z-z_i)^{1/n_i}$ with $n_i$ an integer. We need to distinguish these irregular zeros from the regular ones in $\det M(\vec q)$. For regular zeros $z_r$, $\det M(\vec q) \propto (z-z_r)^{n_r}$ with $n_r$ as positive integer when $z$ close to $z_r$. (The exponent $n_r$ can be larger than unity for the case when the regular zeros can group into multiple ones.) For the case there is one kind of irregular zeros, $\det M(\vec q)$ can be factorized as $\det M(\vec q) = D_r(z)D_i(z)$ with $D_r(z)$ and $D_i(z)$ containing the regular and irregular zeros, respectively. For counting the SSs from the irregular zeros of $\det M(\vec q)$ by performing the contour integral, the contour should encircle each of these irregular zero points $n_i$ times. Under these considerations, the number of the SSs of the energy band is then given by
\begin{eqnarray}
N(k_{\parallel}) = \frac{1}{2m\pi i}\oint_c d\log \det M(\vec q)   \label{ns}
\end{eqnarray}
where the dependence of $k_{\parallel}$ has been written out explicitly, the factor $1/m$ eliminates the multiple counting. A sketch for the contour $c$ is shown in figure 4. We will give an example for the existence of the irregular zeros in Appendix E.

The result given by equation (\ref{ns}) is different from that for the bound states of a central force. In the latter case, the relevant zeros for the bound states are all regular ones ($n_i = 1$) \cite{Levinson,Newton}. The reason is that the energy-momentum relation for free electrons in that case is the simple parabolic form rather than the band structure discussed here. 

The above analysis can be extended to the more general cases of complicated band structures with the maxima/minima not equal valued or with bands overlapping. The presentation is given in Appendix C.

\subsection{Remarks}

Here, we need to make a few remarks in closing this section. 

1. The solution given by equation (\ref{slt}) is obtained by assuming that $A_c$ is invertible. There is the case that the matrix $A_c$ is not invertible at some momenta. The wave functions of the scattering states at these momenta are therefore not uniquely defined. However, the uncertainty of the wave functions at these momenta does not results in any physical consequence as long as the measure of the momentum integral at these momenta to any physical quantity is zero.   

2. The analysis of the SSs is started from equation (\ref{sltx}). As noted above, when the matrix $A_c$ at some momenta is not invertible, the analysis is not valid. In such a case, suppose the parallel momentum is $k_0$. We define the SSs at the parallel momentum $k_0$ as that in the limit $k_{\parallel} \to k_0$. Close to $k_0$, as long as the band structure without abrupt change, the number $m$ and thereby the matrix $A(\vec q)$ do not abruptly change in the limit $k_{\parallel} \to k_0$. The above result for the SSs is still valid. Such a case is discussed with an example given in Appendix E 1. 

3. The SSs are obtained by analytical continuation from the wave function of the scattering states. On the other hand, one may consider a different way to search the SSs instead of doing the analytical continuation. In principle, the wave function of a SS can be expanded in terms of all the $2nL$ evanescent waves. Denote the $2nL$ coefficients of the corresponding waves by a vector $C$ (of dimension $2nL$). According to the same procedure in A, one gets 
\begin{eqnarray}
A(\vec{\tilde q})C = 0, \nonumber
\end{eqnarray}
where $\vec{\tilde q}$ denotes the $2nL$ complex wavenumbers of all the evanescent waves. With the same reason as argued in Appendix B and using Gauss elimination, the matrix $A(\vec{\tilde q})$ can be transformed as
\begin{eqnarray}
RA(\vec{\tilde q}) = M_1 \oplus M_2 \oplus \cdots, \label{gneq}
\end{eqnarray}
where $R$ is a row transformation, and $M_i$ is the $M$ matrix defined above for $i$th band. The rank of $M_i$ is $m_i$. The sum of all $m_i$ equals $2nL$. Note that if there exists overlapping between some bands, those overlapped bands should be considered as a whole as one band. Equation (\ref{gneq}) means that all the solutions to $C$ are determined by 
\begin{eqnarray}
M_ic_i = 0,  ~~{\rm for}~~i = 1, 2,\cdots  \label{gnsl}
\end{eqnarray}
where $c_i$ is a vector of dimension $m_i$ representing the coefficients of the evanescent waves in the $i$th band. Clearly, equation(\ref{gnsl}) is the same equation as given by (\ref{zr}). 

\section{System with TRS}

Here, we apply the previous results to the electron system with the TRS. 

Time-reversal symmetry implies that the electron system is invariant under the operation $\theta = i\sigma_yK$ with $\sigma_y$ the Pauli matrix (operating in the spin space) and $K$ the complex conjugation operator. In real space under consideration, the TRS is reflected by $\theta H(l,k_{\parallel})\theta^{-1}= H(l,-k_{\parallel})$. 

For the bulk states, by the TRS, $\theta H(k)\theta^{-1} = H(-k)$, there exists the Kramers degeneracy between the bulk states $y^{\text{II}}(-k)$ and $\theta y^{\text{I}}(k)$ with energy $E_{\text{II}}(-k) = E_{\text{I}}(k)$. Therefore, the $2n$ energy bands come in pairs. One of the pairs is depicted in figure 5. For the case of only the Kramers degeneracy existing, these eigenstates satisfy the relation \cite{Fu}
\begin{equation}
\theta y^{\text{I}}(k) = y^{\text{II}}(-k) \exp(-i\chi_{-k}),  \label{r1}
\end{equation}
with $\chi_{-k}$ as a phase quantity. There may exist other degeneracy in the states because of other possible symmetries. The TR operator is then represented by an unitary transformation $T$. Suppose the degeneracy in states of each type is $d$. All the energy bands will be grouped to $n/d$ bands. By writing the wave function in a compact form, \begin{eqnarray}
\psi(k)= [y^{\text{I}}_{1}(k),\cdots,y^{\text{I}}_{d}(k),y^{\text{II}}_{1}(k),\cdots,y^{\text{II}}_{d}(k)], \nonumber
\end{eqnarray}
the TR operation on the degenerated states is given by
\begin{eqnarray}
\theta \psi(k) = \psi(-k) T(k),  \label{r2}
\end{eqnarray}
where $T(k)$ is a $2d\times 2d$ unitary matrix. For the case of two type states not degenerated at the same momentum, $T(k)$ is an off-diagonal-block matrix
\begin{eqnarray}
T(k) = \begin{bmatrix}
	0&-t^{t}(-k)\\
	t(k)&0\\
\end{bmatrix}, \label{mtt}  
\end{eqnarray}
where $t(k)$ is a $d\times d$ unitary matrix, and $t^{t}(-k)$ is the transpose of $t(-k)$. For $d = 1$, $t(k)$ is given by $\exp(-i\chi_{-k})$ as defined by equation (\ref{r1}).

\begin{figure}[t]
\centerline{\epsfig{file=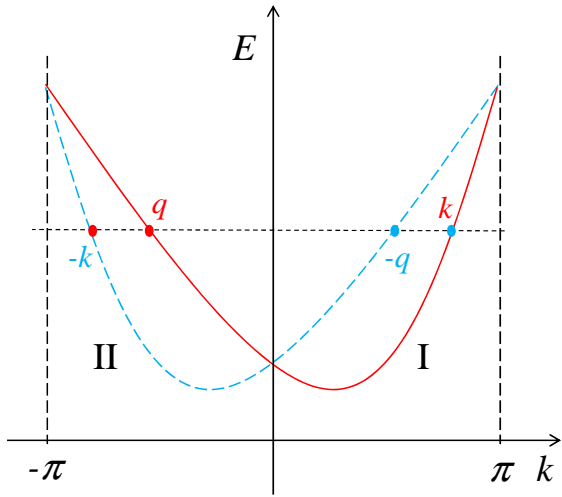,height=7.5cm}}
\caption{A pair of the energy bands I and II with the Kramers degeneracy as functions of momentum. The red (blue) points are the states for defining the wavefunction $\phi_{q,-k}(j)$ [$\phi_{-k,q}(j)$] by equation (\ref{wvph}).} 
\end{figure}

At the TRS points $\Gamma$'s (in the Brillouin zone) that $\theta\psi(\Gamma)= \psi(\Gamma)T(\Gamma)$, $T(\Gamma)$ is an asymmetric matrix because of $\theta^2 = -1$. 

Now, we study the SSs of electrons in the system. For the purpose of proving the correspondence between the change of Kramers degeneracy of the SSs and the bulk time-reversal $Z_2$ invariant, we confine ourselves to the case of $k_{\parallel} = \Gamma_{\parallel}$. For illustrating the physics, here, we study the problem with the simplest case that there is only one maximum/minimum ($m = 1$) in the band of each type as shown in figure 5. We also suppose that there is no any other degeneracy in the two type states except the Kramers degeneracy. This means $d = 1$ here. The discussion for more general cases is given in Appendix D.
 
Since the energy bands of the two type states are overlapped, there are surface reflections between the two bands in the scattering states. Therefore, we need to consider the two overlapped bands as a whole as one band. We then can use the one-band theory given in section 3. The central work for searching the SSs is to find out the zeros of $M({\vec q})$. The starting point is to define the wave function $\phi_{\vec q}(j)$. In the present case, we need to include both states of type I and type II in $\phi_{\vec q}(j)$. Now, it is defined as
\begin{eqnarray}
\phi_{\vec q}(j)\equiv\phi_{q,-k}(j)= [\phi^{\text{I}}_{q}(j), \phi^{\text{II}}_{-k}(j)]. \label{wvph}
\end{eqnarray}
For real wavenumber $q$, when $\phi^{\text{I}}_{q}(j)$ and $\phi^{\text{II}}_{-k}(j)$ are incoming (outgoing) waves, $\phi^{\text{I}}_{k}(j)$ and $\phi^{\text{II}}_{-q}(j)$ are outgoing (incoming) wave. For $\phi^{\text{I}}_{q}(j)$ being an evanescent wave with $q = q_r +iq_i$ and $q_i > 0$, $\phi^{\text{I}}_{k}(j)$ is a growing wave. Since $\phi^{\text{II}}_{-k}(j)$ is obtained by the TR from $\phi^{\text{I}}_{k}(j)$, $-k$ in $\phi^{\text{II}}_{-k}(j)$ is understood as $-k^{\ast}$ and $\phi^{\text{II}}_{-k}(j)$ is a growing wave too. We hereafter write $-k$ as $-k^{\ast}$ explicitly.

With the definition of $\phi_{\vec q}(j)$, the matrix $A(\vec q)$ is then obtained accordingly as
\begin{eqnarray}
A(\vec q) \equiv A(q,-k^{\ast}) = [A^{\text{I}}(q),A^{\text{II}}(-k^{\ast})]. \label{a1}
\end{eqnarray}
By a row transformation $R(q)$, we can obtain
\begin{eqnarray}
R(q) A^{\text{I}}(q)= \underline M^{\text{I}}(q) \label{ra1}
\end{eqnarray}
and 
\begin{eqnarray}
R(q) A(\vec q) = [\underline M^{\text{I}}(q), R(q)A^{\text{II}}(-k^{\ast})]. \label{ra2}
\end{eqnarray}
Here $\underline M^{\text{I}}(q)$ is defined similarly as by equation (\ref{sltb}). The wave function in $M^{\text{I}}(q)$ comes from the type I state. Under the mapping $z = \exp(iq)$, $\det M^{\text{I}}(q)$ as a function of $z$ can be analytically defined within the unit circle in the $z$ plane. [Here, though $M^{\text{I}}(q)$ is a scalar, we keep using the term `matrix' for extending the results to more general case of $m > 1$ and $d > 1$.] The SSs can be analysed accordingly as described in section 3 B. The existence of the SSs is determined by
\begin{eqnarray}
\det M^{\text{I}}(q) = 0. \nonumber
\end{eqnarray}
The solution to the nonvanishing vector $a$ is $a^t = (a^{\text{I}},0)$. According to equation (\ref{ns}), the number of the SSs $\psi^{\text{I}}_b(j;E)$ is obtained by
\begin{eqnarray}
N^{\text{I}}(\Gamma_{\parallel}) = \frac{1}{2\pi i}\oint_{c}d\log \det M^{\text{I}}(q).  \label{ns1}
\end{eqnarray}
Here, for $m$ =1 and $d = 1$, since $\det M^{\text{I}}(q) = M^{\text{I}}(q)$ is an analytical function inside the unit circle in the $z$ plane, the contour $c$ can be defined along the unit circle $|z| = 1$.  
   
On the other hand, we can consider similarly the SSs $\psi^{\text {II}}_b(j;E)$ constructed by the type II evanescent waves. To study the type II SSs, we need to find out the matrix $M^{\text{II}}(q)$. Since $\phi^{\text{I}}_{q}(j)$ and $\phi^{\text{II}}_{-q}(j)$ for real wavenumber $q$ are the bulk states, they are related by the TR transformation. Using this fact, we can obtain a relation between the matrices $A^{\text{I}}(q)$ and $A^{\text{II}}(-q)$ and thereby $M^{\text{I}}(q)$ and $M^{\text{II}}(-q)$. According to this consideration, we derive the relation between the numbers of the two type SSs in follows. 

Under the TR transformation, for real wavenumber $q$, we have
\begin{eqnarray}
\theta \phi^{\text{I}}_{q}(j) = \phi^{\text{II}}_{-q}(j)t(q),  \label{b1}
\end{eqnarray}
which comes from equation (\ref{mtt}). For the matrix $\theta A^{\text{I}}(q)$, we have 
\begin{eqnarray}
\theta A^{\text{I}}(q) = A^{\text{II}}(-q)t(q). \label{b2}
\end{eqnarray}

Note that $i\sigma_y = \sigma_z\sigma_x$ in $\theta$ in equation (\ref{b2}) plays a role of row transformation. In general, the matrix $M(\vec q)$ is composed by the $m$ linearly independent rows of the matrix $A(\vec q)$ as stated in section 3. The action of $i\sigma_y$ on $A(\vec q)$ changes the order of these rows and multiplies some rows by -1. However, after the operation $i\sigma_y$, the original $m$ linearly independent rows are still linearly independent. The further operation $K$ just takes the complex conjugation. As a result, $\det M(\vec q)$ is changed to $\pm\det M^{\ast}(\vec q)$ when $A(\vec q)$ is transformed to $\theta A(\vec q)$. Here, the factor $\pm$ stems from the operation $i\sigma_y$ on $A(\vec q)$.

With the above consideration, from equation (\ref{b2}), we get
\begin{eqnarray}
\pm\det M^{\text{I}\ast}(q) = \det M^{\text{II}}(-q) \det t(q). \label{bb2}
\end{eqnarray}
The number $N^{\text{II}}(\Gamma_{\parallel})$ is then given by
\begin{eqnarray}
N^{\text{II}}(\Gamma_{\parallel}) &=& \frac{1}{2\pi i}\oint_{|z|=1}d\log \det M^{\text{II}}(q) \nonumber\\
&=& \frac{1}{2\pi i}\oint_{|z|=1}d\log [\pm\det M^{\text{I}\ast}(-q)\det t^{-1}(-q)]  \nonumber\\  
&=& N^{\text{I}}(\Gamma_{\parallel})+\frac{1}{2\pi i}\oint_{|z|=1}d\log\det t(q) \nonumber\\  
&=& N^{\text{I}}(\Gamma_{\parallel})+[\log p(\pi)-\log p(-\pi)]/2\pi i,  \label{ns2}
\end{eqnarray}
where the second line is obtained by using the relation (\ref{bb2}), $\det t(q)$ is supposed to be analytical function of $z$ inside the unit circle with all the possible zeros as the regular zeros, and $p(\pm\pi) = \det t(\pm\pi)$ in the last line is the Pfaffian of matrix $T(\Gamma)$ at momentum $\Gamma = (\pm\pi,\Gamma_{\parallel})$.
The number difference $\Delta N(\Gamma_{\parallel})= N^{\text{I}}(\Gamma_{\parallel})-N^{\text{II}}(\Gamma_{\parallel})$ is obtained as
\begin{eqnarray}
\Delta N(\Gamma_{\parallel}) = [\log p(-\pi)-\log p(\pi)]/2\pi i.  \label{nsfr}
\end{eqnarray}
This number should be an integer because of the single valued condition for the wave function. The SSs $\psi^{\text{I}}_b(j;E)$ and $\psi^{\text{II}}_b(j;E)$ are Kramers degenerated if $\Delta N(\Gamma_{\parallel})$ is zero, other wise not degenerated if the number is not zero.

As shown in Appendix D, equation (\ref{nsfr}) is valid for more general cases of the complicated band structures. For a TI, the total number $\Delta N(\Gamma_{\parallel})$ below the Fermi energy, $p(\pm\pi)$ in equation (\ref{nsfr}) is understood as the Pfaffian of matrix $T(\pm\pi)$ operating on the space spanned by the states below the Fermi energy at $q = \pm\pi$. 

Equation (\ref{nsfr}) is the same result as given by Fu and Kane \cite{Fu}. The right hand side of equation (\ref{nsfr}) is obtained as the TR polarization $P_{\theta}$ by Fu and Kane. With $P_{\theta}$, they then define the TR $Z_2$ invariant. Thus, the correspondence between the change of the Kramers degeneracy of SSs and bulk TR $Z_2$ invariant can be proved using the present result. The further proof of this correspondence is the same procedure as given by Fu and Kane \cite{Fu}. We do not repeat it here. 

\section{Conclusion} 

With the generic model, we have studied the eigenstates of the electrons occupying a semi-infinite lattice with a hard surface. We have obtained the exact solution to the wave functions of scattering states. By analytical continuation of the wave functions of scattering states, we obtain the SSs. 

A SS is composed by the evanescent waves of a given band. It is determined by the matrix $M(\vec q)$ at the zeros of $\det M(\vec q)$ in the unit circle in the $z$ plane under the mapping $z=\exp(iq)$. The complex wavenumbers of the evanescent waves are determined by equation (\ref{zrd}). The energy of a SS is given by $E(q)$. The wave function is determined by equation (\ref{bs}). These results mean that all the SSs are classified by the continuous energy bands. In Appendix C, we have generalised the analysis to the cases of all possible complicated band structures. 

The matrix $M(\vec q)$ describes a connection between the wave functions of the bulk states and the SSs. For real wavenumber, the matrix $M(\vec q)$ describes the property of the bulk states. On the other hand, the matrix at the zeros of $\det M(\vec q)$ in the unit circle in the $z$ plane under the mapping $z=\exp(iq)$ determines the SSs. With this connection, we have proved the correspondence between the Kramers degeneracy of the SSs and TR polarization $P_{\theta}$ introduced by Fu and Kane \cite{Fu}. By using this result, the correspondence between of the change of Kramers degeneracy of the SSs and the $Z_2$ invariant of a topological insulator can be proved as done in Ref. \cite{Fu}. 

The analysis is applicable to electron systems in (topological) insulators, superconductors, and semi-metals. We have given some examples in Appendix E for investigating edge states with the theory in graphene with/without spin-orbit interactions and in a $d$-wave superconductor.

A more realistic model for the surface of electron system should include the on-site potentials and the varying hopping within a distance from the surface. The validity of correspondence between the Kramers degeneracy of the SSs and the TR polarization $P_{\theta}$ (introduced by Fu and Kane) can be shown for the realistic model. A paper on the proof is under preparing.

\section*{Acknowledgments}

This work was supported by the National Key R\&D Program of China (2016YFA0202300) and the Robert A. Welch Foundation under Grant No. E-1146.

\appendix
\section{Complex wavenumbers for energy beyond the extrema of the band}
\renewcommand{\theequation}{\thesection\arabic{equation}}
\setcounter{equation}{0}

The wavenumbers $q$'s and $k$'s become pairs of complex conjugates when the energy is beyond the extrema of the band. 

We consider the case of the energy below the lower bound $E_m$ of the band, $E < E_m$. The case of the energy above the upper bound of the band can be analysed accordingly. 

For example, consider the evolution of $k_1$ and $q_2$ in figure 1 in the process of lowering the energy. When $k_1$ becomes complex with negative imaginary part, the wavenumber $q_2$ has positive imaginary part [as stated after equation (\ref{eng})]. By the energy equation $E = E(q_r+iq_i) = E^{\ast}(q_r+iq_i) = E(q_r-iq_i)$, $k_1$ and $q_2$ should be complex conjugates of each other, $k_1 = q_r-iq_i$ and $q_2 = q_r+iq_i$. Further more, by expanding the energy $E(q_r\pm iq_i)$ at $q_m$, we have 
\begin{eqnarray}
E(q_r+iq_i) &=& E(q_m)+\sum_{n=2}^{\infty}\frac{1}{n!}E^{(n)}(q_m)(q_r+iq_i-q_m)^n,   \nonumber\\
E(q_r-iq_i) &=& E(q_m)+\sum_{n=2}^{\infty}\frac{1}{n!}E^{(n)}(q_m)(q_r-iq_i-q_m)^n,   \nonumber\\
E^{(n)}(q_m)&=& \frac{d^nE(q)}{dq^n}|_{q=q_m}. \nonumber
\end{eqnarray}
To satisfy $E(q_r+iq_i) = E(q_r-iq_i)$, we must have $q_r = q_m$ and $E^{(n)}(q_m) =0$ for odd $n$. As a result, we get $k_1 = q_m-iq_i$ and $q_2 = q_m+iq_i$.

\section{Necessary and sufficient condition for the existence of surface states}
\renewcommand{\theequation}{B\arabic{equation}}
\setcounter{equation}{0}

Here, we show equation (\ref{zrd}) is the necessary and sufficient condition for the existence of surface states. Note that equation (\ref{mtbc}) can be rewritten as 
\begin{eqnarray}
[A(\vec q),-A(\vec{\tilde k})] Z = A(\vec k) S, \label{apb1}
\end{eqnarray}
with
\begin{eqnarray}
Z &=&\begin{pmatrix}
	I \\
	D \\
\end{pmatrix},                         \nonumber  
\end{eqnarray}
where $I$ is an unit matrix of dimension $m\times m$. The ranks of $A(\vec q)$ and $A(\vec{\tilde k})$ are $m$ and $2nL-m$ (see the statement in Sec. III A), respectively. By row transformation $R(\vec q)$, the matrix $[A(\vec q),-A(\vec{\tilde k})]$ is transformed as 
\begin{eqnarray}
R(\vec q)[A(\vec q),-A(\vec{\tilde k})]=\begin{bmatrix}
	M(\vec q)& C_1\\
	0&C_2\\
\end{bmatrix},  
\nonumber
\end{eqnarray}
where $C_1$ and $C_2$ are matrices of dimensions $m\times (2nL-m)$ and $(2nL-m)\times (2nL-m)$, respectively. Since the rank of the matrix $[A(\vec q),-A(\vec{\tilde k})]$ is $2nL$, the rank of $C_2$ is $2nL-m$ [the same rank of $A(\vec{\tilde k})$]. Therefore, the $2nL-m$ rows of $C_2$ are linear independent vectors. All the rows in $C_1$ are linear dependent on these $2nL-m$ rows of $C_2$. By a row transformation $R_c$, the matrix $C_1$ can be eliminated. We then get
\begin{eqnarray}
R_cR(\vec q)[A(\vec q),-A(\vec{\tilde k})]=\begin{bmatrix}
	M(\vec q)& 0\\
	0&C_2\\
\end{bmatrix}.  
\label{apb2}
\end{eqnarray}
The row transformations on $A(\vec q)$ and $A(\vec{\tilde k})$ amount to choose the linear independent row vectors and eliminate the linear dependent ones. Then, equation (\ref{apb1}) can be written as
\begin{eqnarray}
\begin{bmatrix}
	M(\vec q)& 0\\
	0&C_2\\
\end{bmatrix}\begin{pmatrix}
	I \\
	D \\
\end{pmatrix} =  R_cR(\vec q)A(\vec k)S. 
\label{apb3}
\end{eqnarray}

Now, consider the analytical continuation by changing the energy beyond the extrema of the band. At certain energy $E$, suppose there exists a non-zero vector $a$ that $Sa = 0$. By multiplying equation (\ref{apb3}) with $a$ by right-product, we have
\begin{eqnarray}
\begin{bmatrix}
	M(\vec q)& 0\\
	0&C_2\\
\end{bmatrix}\begin{bmatrix}
	a\\
	Da\\
\end{bmatrix} = 0. 
\label{apb4}
\end{eqnarray}
From equation (\ref{apb4}), we get $M(\vec q)a = 0$. The second equation $C_2Da = 0$ is then automatically satisfied because of $Da = 0$ [see equation (\ref{sltx})]. For the existence of the non-zero vector $a$, the necessary and sufficient condition is $\det M(\vec q) = 0$.

\section{General Band Structures}
\renewcommand{\theequation}{C\arabic{equation}}
\setcounter{equation}{0}

In real problems, the energy band structures are generally not the simple one as that discussed in section 3. In general, there is an equal number of maxima and ninima in an band. However, all maxima (minima) may not have the same energy. Here, we illustrate how to apply the theory in section 3 to general cases with more complicated band structures. 

(i) {\it Band with non-equal-valued maxima} ({\it minima}). Without loss of generality, we consider the case that there is one maximum with a different value from the highest maximum. Such an energy band with $m = 2$ is depicted in figure C1. Above a critical energy $E_c$, the number of the incoming waves (also of the outgoing waves) is $m-1$. The absent incoming and outgoing waves are lost at the wavenumber $q_{c0}$. As discussed in Appendix A, the wavenumbers of the absent incoming and outgoing waves for $E_c < E < E_M$ have become $q_{c0}+iq_i$ and $q_{c0}-iq_i$, respectively. As depicted in top panel in figure C1, we use the red-dashed vertical line to present the absent evanescent wave with wavenumber $q_{c0}+iq_i$. Under the mapping $z = \exp(iq)$, this vertical line is mapped to the line along $o-q_{c0}$ direction with angle equal to $q_{c0}$. 

\begin{figure}[t]
\centerline{\epsfig{file=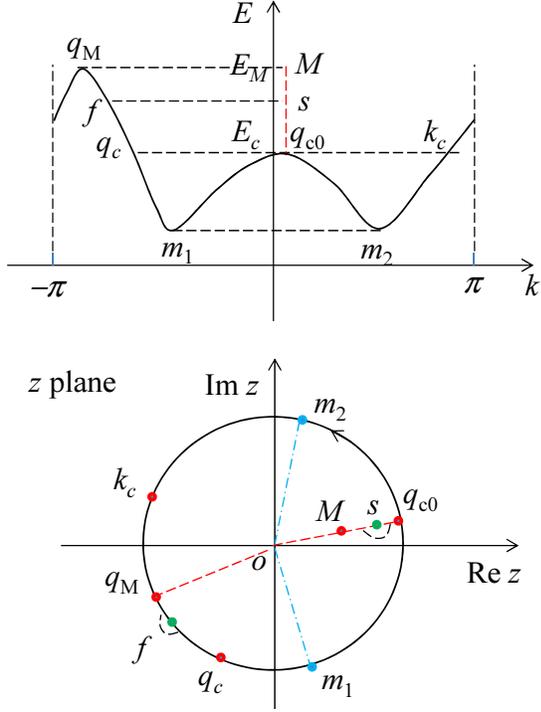,width=8.cm}}
\caption{Top panel: Sketch of an energy band. The numbers of real incoming (outgoing) waves above and below $E_c$ are different. The red vertical line represents the evanescent wave. The point $s$ denotes a possible in-band SS. Lower panel: $z$ plane for the mapping $z = \exp(iq)$. The points on the energy curve are mapped to the corresponding points in the $z$ plane. The red dashed and the blue dash-dot lines have the same meanings as in figure 3. The green points $f$ and $s$ are the possible zeros of $\det M(\vec q)$. Close to the point $f$, the contour for the integral in equation (\ref{ns}) is along the small dashed arc.} 
\end{figure}

For the scattering states of energy $E$ with $E_c < E < E_M$, we need to add the evanescent wave to $\varphi$ and to drop the growing wave from $\phi_{\vec k}$ since it does not satisfy the boundary condition at $j \to \infty$. We then get a similar expression for the wave function as given by equation (\ref{cwv}). 

{\it In-band surface states.} On the other hand, there may be the possibility that the evanescent wave evolves to a SS. This is an in-band SS. To get the criterion for the existence of this SS, we still consider the matrix $M(\vec q)$ of dimension $m\times m$ for $E<E_c$. Now, one column of $M(\vec q)$, say the $m$th column, is associated with the evanescent wave when $E_c<E<E_M$. Suppose by row transformation $M(\vec q)$ is transformed to 
\begin{eqnarray}
M(\vec q) \to \begin{bmatrix}
	 M_{(m-1)\times(m-1)}&x_{(m-1)\times 1}\\
	0_{1\times(m-1)}&r\\
\end{bmatrix}.               
\end{eqnarray}
The block $M_{(m-1)\times(m-1)}$ is associated with the $(m-1)$ real (incoming or outgoing) waves and thereby $\det M_{(m-1)\times(m-1)} \ne 0$. Under analytical continuation as $E$ going above $E_c$ from below, if $r = 0$ at a certain energy ($< E_M$), we will get an nonzero vector $a^{t} = (0,\cdots,0,1)$ (with the superscript $t$ denoting the transpose) satisfying $M(\vec q)a = 0$ and thus obtain a SS coming from the evanescent wave. Clearly, the criterion for the existence of this SS is still given by equation (\ref{zrd}).

As shown in figure C1, for energy $E$, $E_c < E < E_M$, the real incoming and outgoing waves are defined respectively on the arcs $q_M-q_c$ and $k_c-q_M$ on the unit circle in the $z$ plane, while the evanescent wave is defined on the line $q_{c0}-M$ along the radial direction. When the real wavenumber $q$ varies from $k_{c} \to q_M \to q_{c}$, the energy varies as $E_c \to E_M \to E_c$. Meanwhile, the point reflecting the evanescent wave moves as $q_{c0} \to M \to q_{c0}$ in the $z$ plane. The green points $f$ and $s$ are the zeros of $\det M(\vec q)$ giving rise to the single in-band SS. 

The contour of the integral in equation (\ref{ns}) is now deformed with a small arc close to point $f$ shown in lower panel of figure C1. The green points are the possible zeros of $\det M(\vec q)$. Instead to consider the small arc, the point at $f$ can be considered as close to the unit circle from the inside. It means that the wavenumber at point $f$ has an infinitesimal small positive imaginary part. With the small arc close to the point $f$ in the contour, all the zeros for the in-band state are included in the integral. Since the evanescent wave defined on the line $q_{c0}-M$ and the real wave defined in the arc $k_c-q_M-q_c$ are associated with the same energy, the small arc close to point $s$ is a correspondence to the one close to $f$. 

For the energy $E$ above the upper bound $E_M$ of the band, all the waves involved in composing the matrix $M(\vec q)$ become evanescent waves and the more possible zero points are placed on the radial lines $o-q_M$ and $o-M$. Here, the angle of radial line $o-q_M$ is the momentum $q_M$.

The analysis can be extended to the band structure with several non-equal maxima. For a more general band structure, all the minima may not equal valued either. We can extend the above analysis to this case as well. 

There may be the cases that the highest minimum energy is larger than the lowest maximum energy. An in-band SS may consist of the evanescent waves defined in both the red dash and the blue dash-dot lines in figure C1. These lines in the lower panel of figure C1 have the similar meanings as defined in figure 3. 

{\it Zeros distribution.} From the example and the analysis in section 3 for the zeros of $\det M(\vec q)$, we get the following conclusions. (1) For the in-band SSs, some of the zeros of $\det M(\vec q)$ distribute on the dashed and/or dash-dot lines in the unit circle and the rest zeros distribute on the unit circle as shown in figure C1. (2) For the SSs with energy above the top bound of the band, all the zeros distribute on the red dashed lines. (3) For the SSs with energy below the lower bound of the band, all the zeros distribute on the blue dash-dot lines. At these zeros of $\det M(\vec q)$, the energies are real and the wave functions represent physical SSs. 

(ii) {\it Overlapped bands.} When there exists overlapping between energy bands, an incoming wave can be reflected to outgoing waves of all overlapped bands. All the overlapped bands should be considered as one band. The upper bound of the band is given by the highest maximum in the overlapped bands, while the lower bound is the lowest minimum. The rank $m$ is then the sum of the ranks of the individual bands. An example of two overlapped bands is shown in figure C2. The vertical lines for defining the evanescent waves are also depicted in figure C2. Now, the incoming (outgoing) waves include all that of the individual bands. Thus, the formalism given in section 3 along with the extension discussed in case (i) is applicable to the present case. 

\begin{figure}[t]
\centerline{\epsfig{file=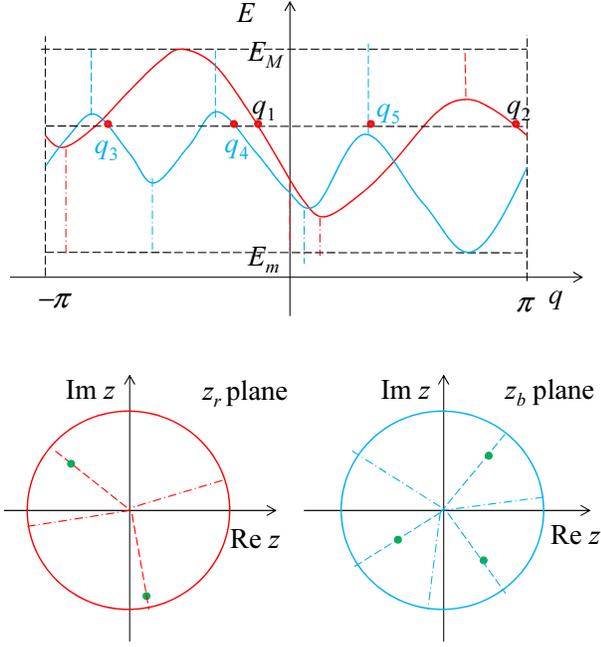,width=8.5cm}}
\caption{Top panel: Two energy bands with overlapping. The red and blue vertical dashed (dash-dot) lines are the energy lines above the maxima (below the minima) for the evanescent waves with complex wavenumbers. The red points denote the waves at the same energy for defining $\phi_{\vec q}(j)$. Lower panel: The $z_r$-plane and $z_b$-plane corresponding to the red and blue energy bands in the top panel. The radial lines inside the unit circle correspond to the lines in the top panel. The green points on the dashed lines denote the possible zeros for energy above the upper energy band.}
\end{figure}

To apply the contour integral given by equation (\ref{ns}) for the present case, here, we need to clearly define the $z$ mapping and the contour in the $z$ plane for energy bands with overlapping. 

{\it The $z$ mapping.} We take $z = z_{r} = \exp(iq_{r1})$ of the red band shown in figure C2. Then, all other $q_{\mu}$'s of red band and $q_{\nu}$'s of blue band in the incoming (outgoing) waves are determined by $E_r(q_{\mu}) = E_b(q_{\nu}) = E_r(q_{r1})$. We can also take $z_b = \exp(iq_{b1})$, which is a function of $z$. Under the mapping, the red and blue bands are mapped to $z_r$ plane and $z_b$ plane, respectively. In figure C2, each dashed (dash-dot) line is for defining the evanescent wave for energy above (below) the maximum (minimum). Denote the numbers of maxima of red band as $m_r$ and the blue band as $m_b$. Their least common multiple is $m_l$. Now, the closed contour is defined as $z$ running $m_l/m_r$ times on the contour $c_r$ in the $z_r$ plane. With this definition, $z_b$ runs $m_l/m_b$ times on the contour $c_b$ in the $z_b$ plane.

As shown in figure C2, corresponding to one SS, the zeros of $\det M(\vec q)$ are the $m_r$ and $m_b$ points in the $z_r$ plane and $z_b$ plane, respectively. We need to perform the contour integral in one plane, for example, the $z_r$ plane. Since $z_b$ is a function of $z_r$, the zeros in $z_b$ plane are automatically counted. In the $z_r$ plane, the contour encircles each zero $m_l/m_r$ times. We define this number as the encircling number for each zero. The total encircling number in the $z_r$ plane is $m_r\cdot m_l/m_r = m_l$. Similarly, the total encircling number in the $z_b$ plane is $m_b\cdot m_l/m_b = m_l$. We need to divide the contour integral by $m_l$. 

The analysis can be extended to the case of more than two overlapped bands.  

(iii) {\it Degenerated bands.} For an energy band with degeneracy $d$, there exist surface reflections between the degenerated states. This is a special case of $d$ bands with overlapping. The wave functions for the scattering states can be still written as equation (\ref{cwv}) but with the components of $\phi_{\vec q}(j)$ understood as
\begin{eqnarray}
\phi_{q_{\mu}}(j) &=& [\phi_{q_{\mu}1}(j),\cdots,\phi_{q_{\mu}d}(j)], \nonumber 
\end{eqnarray}
Though the momenta of the components of $\phi_{q_{\mu}}(j)$ are the same $q_{\mu}$, they are independent column vectors. By definition, the rank of $M(q)$ now is $md$. On the other hand, for counting the number of SSs, we note that the contour for integral is defined in the single $z$ plane. For one SS, the number of zeros of $\det M(q)$ is $m$. Therefore, the formula for counting the number of SSs is the same as equation (\ref{ns}). 

So far, we have considered all possible cases of the band structures in real problems.

\section{SS in sytems with TRS}
\renewcommand{\theequation}{D\arabic{equation}}
\setcounter{equation}{0}

Here, we analyse the eigenstates of electrons with TRS for complicated band structures. The number $m$ of the maxima/minima of the band can be larger than unity. There may be other degeneracy in the two type states besides the Kramers degeneracy. For these more general cases, we derive equation (\ref{nsfr}).

(i) The two type states not degenerated at the same momentum. A sketch of the energy bands is given in figure D1. For the scattering states, there are reflections between the two type states. The wave functions of scattering states can be written as
\begin{eqnarray}
\psi^{\text{I}}(j;E) &=& \phi^{\text{I}}_{\vec q}(j)-\phi^{\text{I}}_{\vec k}(j)S^{\text{I,1}}  
-\phi^{\text{II}}_{-\vec q^{\ast}}(j)S^{\text{I,2}}\nonumber\\
&& -\varphi^{\text{I}}_{\vec{\tilde k}}(j)D^{\text{I,1}}-\varphi^{\text{II}}_{-{\vec{\tilde k}}^{\ast}}(j)D^{\text{I,2}}  \nonumber\\
\psi^{\text{II}}(j;E) &=& \phi^{\text{II}}_{-\vec k^{\ast}}(j)-\phi^{\text{I}}_{\vec k}(j)S^{\text{II,1}}  
-\phi^{\text{II}}_{-\vec q^{\ast}}(j)S^{\text{II,2}}   \nonumber \\
&& -\varphi^{\text{I}}_{\vec{\tilde k}}(j)D^{\text{II,1}}-\varphi^{\text{II}}_{-{\vec{\tilde k}}^{\ast}}(j)D^{\text{II,2}}  \nonumber 
\end{eqnarray}
where $S$'s and $D$'s are matrices.

According to the arguments given in Appendix C, the two bands should be treated as one band. The wave functions of scattering states can be written in the compact form as given by equation (\ref{cwv})  
with the waves understood as
\begin{eqnarray}
\phi_{\vec q}(j) &\equiv& \phi_{\vec q,-\vec k^{\ast}}(j) = [\phi^{\text{I}}_{\vec q}(j), \phi^{\text{II}}_{-\vec k^{\ast}}(j)],
\label{wvct} \\
\phi_{\vec k}(j) &\equiv& \phi_{\vec k,-\vec q^{\ast}}(j) = [\phi^{\text{I}}_{\vec k}(j), \phi^{\text{II}}_{-\vec q^{\ast}}(j)],
\nonumber\\
\varphi_{\vec{\tilde k}}(j) &\equiv& [\phi^{\text{I}}_{\vec{\tilde k}}(j), \phi^{\text{II}}_{-{\vec{\tilde k}}^{\ast}}(j)]. \nonumber
\end{eqnarray}

Under the TR operation, $\phi_{\vec q,-\vec k^{\ast}}(j)$ transforms as
\begin{eqnarray}
\theta \phi_{\vec q,-\vec k^{\ast}}(j) = \phi_{\vec k,-\vec q^{\ast}}X(\vec q,\vec k),  \label{eqb1}
\end{eqnarray}
with matrix $X(\vec q,\vec k)$ defined as 
\begin{eqnarray}
X(\vec q,\vec k) = \begin{bmatrix}
	0&-t^{t}(\vec k)\\
	t(\vec q)&0\\
\end{bmatrix}, \label{mtxx}  
\end{eqnarray}
which comes from equation (\ref{mtt}). Here, $t(\vec q)$ is a diagonal (block) matrix defined by
\begin{eqnarray}
t(\vec q) = diag[t(q_1),\cdots,t(q_m)].   \nonumber
\end{eqnarray}
For the case there exist degeneracy other than the Kramers degeneracy in the states of each type, $t(q)$ is a matrix as that appeared in equation (\ref{mtt}). 

{\it Remark.} In equation (\ref{mtxx}), some of the wavenumbers may be complex. Suppose there is a maximum at momentum $q_{c0}$ less than the largest maximum in the energy band of type I states. Such a structure has been shown in figure C1. Then, there is a component with complex wavenumber $q_{c0} + iq_i$ (with $q_i > 0$) in $\vec q$. We need to define the matrix $t(q)$ for complex $q$. As seen from the example shown in figure C1, when the evanescent wave goes from $q_{c0}$ to $M$ and then returns to $q_{c0}$ on the same way, the matrix $t(q)$ at $q=q_{c0}$ should not change. Therefore, we can define $t(q) = t(q_{c0})$ for the evanescent wave running on its path. For the a growing wave with $q_{c0} - iq_i$, we can similarly define $t(q) = t(q_{c0})$.

Corresponding to equation (\ref{wvct}), the matrix $A(\vec q)$ is now defined as 
\begin{eqnarray}
A(\vec q) \equiv A(\vec q,-\vec k^{\ast}) = [A^{\text{I}}(\vec q),A^{\text{II}}(-\vec k^{\ast})]. \label{eqa1}
\end{eqnarray}
Under the TR operation, $A(\vec q)$ transforms as 
\begin{eqnarray}
\theta A(\vec q, -\vec k^{\ast}) = A(\vec k,-\vec q^{\ast})X(\vec q,\vec k). \label{eqb2}
\end{eqnarray}

\begin{figure}[t]
\centerline{\epsfig{file=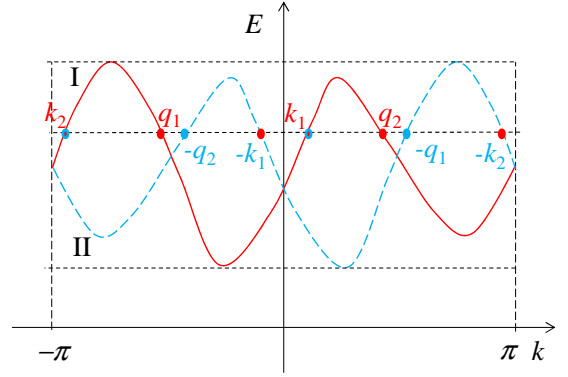,width=8.cm}}
\caption{Sketch of energy bands for type I (red) and type II (blue) states. The red and blue points denote the components including both type I and II states in $\phi_{\vec q}(j)$ $\phi_{\vec k}(j)$, respectively.}
\end{figure}
  
As analysed in section 4, when the components in $\phi^{\text{I}}_{\vec q}(j)$ present evanescent (growing) waves, the components in$\phi^{\text{II}}_{-\vec k^{\ast}}(j)$ are growing (evanescent waves). In the present case, since the maxima (minima) may not equal valued, only some of the momenta in $\vec q$ and in $-\vec k^{\ast}$ become complex when the energy between the maxima (minima) of a given band. When the energy beyond the extrema, all the momenta become complex. The in-band SSs and the in-gap SSs can be analysed equally in follows with the help as described in Appendix C. Allowing the components (or some of them) in $\phi^{\text{I}}_{\vec q}(j)$ becoming the evanescent waves when changing the energy, we then construct the matrix $A^{\text{I}}({\vec q})$ and get the matrix $M^{\text{I}}({\vec q})$ by row transformation $R(\vec q)$ on $A^{\text{I}}({\vec q})$. Under the mapping $z = \exp(iq_1)$, the matrix $M^{\text{I}}(\vec q)$ as a function of $z$ can be analytically defined inside the unit circle in the $z$ plane. The existence of SSs is determined by 
\begin{eqnarray}
\det M^{\text{I}}(\vec q) = 0.  \nonumber 
\end{eqnarray}
A nonzero vector $a^{\text{I}}$ is then determined by
\begin{eqnarray}
M^{\text{I}}(\vec q) a^{\text{I}}= 0.  \nonumber 
\end{eqnarray}
From $a^{\text{I}}$, we obtain a nonzero vector $a^t = (a^{\text{I}t},0)$. The vector $a$ satisfies the equation
\begin{eqnarray}
R(\vec q)A(\vec q,\vec k) a= 0.  \nonumber 
\end{eqnarray}
The SSs so obtained consists of only the evanescent waves of type I states. The number of the SSs is given by 
\begin{eqnarray}
N^{\text{I}}(\Gamma_{\parallel}) = \frac{1}{2\pi i}\oint_{c}d\log \det M^{\text{I}}(\vec q).  \nonumber
\end{eqnarray}

Similarly, we can determine the another type SSs $\psi^{\text{II}}_b(j;E)$ by the corresponding matrix $M^{\text{II}}({\vec q})$. As stated in section 4, $\det M^{\text{I}}(\vec q)$ is changed to $\pm\det M^{\text{I}\ast}(\vec q)$ when $A^{\text{I}}({\vec q})$ is transformed to $\theta A^{\text{I}}({\vec q})$. From equations (\ref{mtxx}) and (\ref{eqb2}), we get 
\begin{eqnarray}
\pm\det M^{\text{I}\ast}(\vec q) = \det M^{\text{II}}(-\vec q^{\ast})\det t(\vec q). \label{eqb3}
\end{eqnarray}
The number of the type II SSs is calculated by
\begin{eqnarray}
N^{\text{II}}(\Gamma_{\parallel}) &=& \frac{1}{2m\pi i}\oint_{c}d\log \det M^{\text{II}}(\vec q)  \nonumber\\ 
&=& \frac{1}{2m\pi i}\oint_{c}d\log [\pm\det M^{\text{I}\ast}(-\vec q^{\ast})\det t^{-1}(-\vec q^{\ast})]  \nonumber\\  
&=& N^{\text{I}}(\Gamma_{\parallel})+\frac{1}{2m\pi i}\oint_{c}d\log [\det t(\vec q)]   \label{nsd1}
\end{eqnarray}
where the second line is obtained by using equation (\ref{eqb3}). Here, for the contour integral of a function $f(q)$ analytically defined inside the unit circle in the $z$ plane under the mapping $z= \exp(iq)$, we have 
\begin{eqnarray}
\oint_{c}d\log f^{\ast}(-q^{\ast}) &=& [\oint_{c}d\log f(-q^{\ast})]^{\ast}   \nonumber\\
&=& -[\oint_{c}d\log f(q)]^{\ast}   \nonumber\\
 &=& \oint_{c}d\log f(q).   \label{cir}
\end{eqnarray}
We have used this relation for getting the last line of equation (\ref{nsd1}). 

By supposing the matrix $t(\vec q)$ to be analytical function of $z$ inside the unit circle with all the possible zeros as the regular ones, the contour $c$ in the last line of equation (\ref{nsd1}) can be redefined on the unit circle in the $z$ plane. Note that $\det t(\vec q)=\prod_{\mu} t(q_{\mu})$. For a real $q_1$, some of other momenta $q_{\mu}$ may be complex. Without loss of generality, we suppose that there is one of them is complex as shown in figure C1. For performing the remained integral, we include the path $q_{c0} \to M \to q_{c0}$ as shown in the lower panel figure C1 in the contour. The contribution from this path to the contour integral is zero. However, by including this path, we see that when the variable $z = \exp(iq_1)$ runs a round on the total contour all other variables $z_{\mu} = \exp(iq_{\mu})$ finish equally their round running on the total contour. With this consideration, the contour integral in equation (\ref{nsd1}) reduces to all the $m$ equal individual contour integrals. For the number difference between two type SSs, we get
\begin{eqnarray}
\Delta N(\Gamma_{\parallel}) &\equiv& N^{\text{I}}(\Gamma_{\parallel})-N^{\text{II}}(\Gamma_{\parallel}) \nonumber\\
&=& -\frac{1}{2\pi i}\oint_{c}d\log [\det t(q)]   \nonumber\\
&=& [\log\det t(-\pi)-\log\det t(\pi)]/2\pi i,   \label{nsrf2}
\end{eqnarray}
where $p(\pm\pi) = \det t(\pm\pi)$ is the Pfaffian of matrix $T(\Gamma)$ at momentum $\Gamma = (\pm\pi,\Gamma_{\parallel})$.
 
(ii) Type I and type II states degenerated at the same momentum. In this case, the two type states cannot be uniquely distinguished. The total degeneracy at a momentum is $2d$. The component of $\phi_{\vec q}(j)$ at $q_{\mu}$ now is the one with $2d$ components, 
\begin{eqnarray}
\phi_{q_{\mu}}(j)= [\phi_{\mu 1}(j),\cdots,\phi_{\mu d}(j),\phi_{\mu d+1}(j),\cdots,\phi_{\mu 2d}(j)]. \nonumber 
\end{eqnarray}
The band structure in this case can be considered as the red and blue curves in figure D1 merge into one curve under continuous change. Therefore, for energy beyond the extrema, when half of the components in $\phi_{q_{\mu}}(j)$ are evanescent waves with wavenumber $q^+_{\mu}$, the another half are growing waves with wavenumber $q^-_{\mu}$. For reflecting this fact, we denote the wave $\phi_{q_{\mu}}(j)$ as $\phi_{q^+_{\mu},q^-_{\mu}}(j)$. Certainly, the wavenumbers $q^+_{\mu}$ and $q^-_{\mu}$ can be the same real wavenumber for energy within the band. Since the two type states are not uniquely distinguished, the order of the components in $\phi_{q_{\mu}}(j)$ are not necessarily arranged as the first $d$ components for the waves $q^+_{\mu}$ and the rest for the waves $q^-_{\mu}$. However, in principle, $\phi_{q^+_{\mu},q^-_{\mu}}(j)$ can be arranged in such an order. Hereafter, we consider $\phi_{q^+_{\mu},q^-_{\mu}}(j)$ in this order and write it as
\begin{eqnarray}
\phi_{q^+_{\mu},q^-_{\mu}}(j) = [\phi_{q^+_{\mu}}(j), \phi_{q^-_{\mu}}(j)].
\end{eqnarray}

With the definition of $\phi_{q^+_{\mu},q^-_{\mu}}(j)$, we then can construct matrix $A(\vec q^+,\vec q^-)$,
\begin{eqnarray}
A(\vec q^+,\vec q^-) = [A(\vec q^+), A(\vec q^-)].
\end{eqnarray}
Now, the SSs can be analysed in parallel as that in (i). By a row transformation $R(\vec q^+)$, we obtain the matrix $M_1(\vec q^+)$ from $A(\vec q^+)$. Under the mapping $z = \exp(iq_1)$, the matrix $M_1(\vec q^+)$ as a function of $z$ can be analytically defined inside the unit circle in the $z$ plane. The SSs is determined by 
the matrix $M_1(\vec q^+)$. The number of the SSs so obtained is given by 
\begin{eqnarray}
N_1(\Gamma_{\parallel}) = \frac{1}{2\pi i}\oint_{c}d\log \det M_1(\vec q).  \nonumber
\end{eqnarray}

On the other band, there is a solution with $\phi_{q_{\mu}}(j) = \phi_{q^-_{\mu},q^+_{\mu}}(j)$ for energy beyond the extrema. One then defines the matrix $A(\vec q^-,\vec q^+)$ and then obtains a matrix $\det M_2(\vec q^+)$ from $A(\vec q^-,\vec q^+)$ by row transformation. Notice that the waves $\phi_{q^+,q^-}(j)$ and $\phi_{-q^+,-q^-}(j)$ are related by the TR operation, 
\begin{eqnarray}
\theta \phi_{q^+,q^-}(j) = \phi_{-q^+,-q^-}(j)T(q),  \label{trs1}
\end{eqnarray}
with
\begin{eqnarray}
T(q) = \begin{bmatrix}
	0&-t^{t}(-q)\\
	t(q)&0\\
\end{bmatrix}, \nonumber  
\end{eqnarray}
where we have used the fact that for complex wave numbers $q^+$ and $q^-$ the transformation is defined as at the corresponding real wavenumber $q$ as indicated in (i). With equation (\ref{trs1}), we get
\begin{eqnarray}
\theta A(\vec q^+,\vec q^-) &=& A(-\vec q^+,-\vec q^-) T(\vec q),  \label{eqd1}\\
T(\vec q) &=& diag[T(q_1),\cdots,T(q_m)].   \label{mtx1} 
\end{eqnarray}

From equation (\ref{eqd1}), we get the relation 
\begin{eqnarray}
\pm\det M_1^{\ast}(\vec q^+) = \det M_2(-\vec q^-) \det t(\vec q).  \label{det12}
\end{eqnarray}
The number of the SSs determined by the matrix $M_2(\vec q^+)$ is given by
\begin{eqnarray}
N_2(\Gamma_{\parallel}) &=& \frac{1}{2\pi i}\oint_{c}d\log \det M_2(\vec q)  \nonumber\\
&=& \frac{1}{2m\pi i}\oint_{c}d\log [\pm\det M_1^{\ast}(-\vec q^{\ast}) \det t^{-1}(-\vec q^{\ast})]  \nonumber\\
&=& N_1(\Gamma_{\parallel})+\frac{1}{2m\pi i}\oint_{c}d\log\det t(\vec q)  \nonumber\\
&=& N_1(\Gamma_{\parallel})+[\log P(\pi)-\log P(-\pi)]/2\pi i   \label{nst2} 
\end{eqnarray}
where we have used equations (\ref{det12}) and (\ref{cir}), $\det t(\vec q)$ is supposed as analytical function of $z$ inside the unit circle with all the possible zeros as the regular ones, and $P(\pm\pi) = \det t(\pm\pi)$ is the Pfaffian of matrix $T(\pm\pi)$. 

From equation (\ref{nst2}), we obtain the number difference $\Delta N(\Gamma_{\parallel})=N_1(\Gamma_{\parallel})-N_2(\Gamma_{\parallel})$ as
\begin{eqnarray}
\Delta N(\Gamma_{\parallel}) &=& [\log P(-\pi)-\log P(\pi)]/2\pi i.         \label{eqd2}
\end{eqnarray}
Note that $P(\pm\pi)$ reduces to $p(\pm\pi)$ when the two type states are not degenerated at the same momentum. Therefore, equation (\ref{eqd2}) gives rise to the general results including the case (i).

For the case of more bands with overlapping, the problem can be treated similarly as discussed in section 4.2.

Finally, for a TI, the total number $\Delta N(\Gamma_{\parallel})$ below the Fermi energy is obtained by summing all the contributions from the bands below the Fermi energy. Then $P(\pm\pi)$ in equation (\ref{eqd2}) is understood as the Pfaffian of matrix $T(\pm\pi)$ operating on the space spanned by the states below the Fermi energy at $q = \pm\pi$. 

\section{Examples}
\renewcommand{\theequation}{E\arabic{equation}}
\setcounter{equation}{0}

\subsection{Edge states of electrons with SOI in graphene} 

In this subsection, we apply the present theory to study the edge states of electrons with spin-orbit interactions in graphene for two simple cases. The Hamiltonian is given by \cite{Fu2,Fu3}
\begin{eqnarray}
H = t\sum_{\langle ij\rangle}C^{\dagger}_iC_j+i\lambda\sum_{\langle\langle ij\rangle\rangle}\nu_{ij}C^{\dagger}_i\sigma_zC_j  \label{hm}
\end{eqnarray}
where $C^{\dagger}_i = (c^{\dagger}_{i\uparrow},c^{\dagger}_{i\downarrow})$ with $c^{\dagger}_{is}$ creating an electron of spin $s$ on site $i$, $\lambda$ is the spin-orbit interaction strength, $\langle ij\rangle$-sum runs over the nearest-neighbor sites, $\langle\langle ij\rangle\rangle$-sum runs over the next nearest neighbor sites, $\sigma_z$ is the Pauli matrix operating on the spin space, and $\nu_{ij}$ is defined as  
\begin{eqnarray}
\nu_{ij}=(\vec d_1\times\vec d_2)\cdot \hat z/|\vec d_1\times\vec d_2|  \label{nu}
\end{eqnarray}
with $\vec d_1$ and $\vec d_2$ as the consecutive vectors when hopping from site $j$ to site $i$ as shown in figure E1, and $\hat z$ the unit vector in normal direction. We will use the units in which $t = 1$.

\begin{figure}[t]
\centerline{\epsfig{file=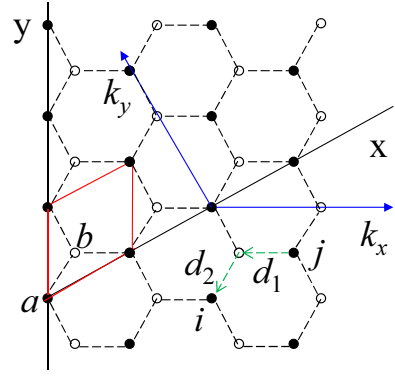,width=6.cm}}
\caption{Graphene lattice with a zigzag edge. The red diamond is a unit cell containing atoms $a$ and $b$. The axes of real $(x,y)$ and reciprocal $(k_x,k_y)$ spaces are also shown. The vectors $d_1$ and $d_2$ show the route of an electron hopping from site $j$ to site $i$.}
\end{figure}

For a semi-infinite honeycomb lattice with a zigzag edge, the coordinates in real and the reciprocal space are shown in figure E1. After the Fourier transform along the edge direction, the Hamiltonian reads
\begin{eqnarray}
H = \sum_{ij k_y}C^{\dagger}_i(k_y)H_{ij}(k_y)C_j(k_y) \label{thm}
\end{eqnarray}
where $k_y$ is the momentum parallel to the edge. Here, $H_{ij}(k_y)$'s are given by 
\begin{eqnarray}
H_{jj}(k_y) &\equiv& H(0) \nonumber\\
&=&(1+\cos k_y)\tau_x +\sin k_y \tau_y +2\lambda\sin k_y \tau_z\sigma_z \nonumber\\
H_{j,j+1}(k_y) &\equiv& H(1) \nonumber\\
&=&\tau_-  +i\lambda[1-\exp(-ik_y)] \tau_z\sigma_z \nonumber\\
H_{j,j-1}(k_y) &\equiv& H(-1)\nonumber\\
&=&\tau_+  -i\lambda[1-\exp(ik_y)] \tau_z\sigma_z \nonumber 
\end{eqnarray}
where the Pauli matrices $\tau$'s operate on the sublattice space $(a,b)$, $C^{\dagger}_i(k_y) = (c^{\dagger}_{ia\uparrow},c^{\dagger}_{ib\uparrow},c^{\dagger}_{ia\downarrow},c^{\dagger}_{ib\downarrow})$ with $c^{\dagger}_{ia\sigma}$ creating an electron of spin $\sigma$ on atom $a$ of unit cell $i$, and the $ij$-sum runs over the $x$-coordinates of unit cells. For the bulk states, the Hamiltonian $H(k)$ is given by
\begin{eqnarray}
H(k) &=& d_1(k)\tau_x\sigma_0 +d_2(k)\tau_y\sigma_0 +d_3(k)\tau_z\sigma_z, \nonumber 
\end{eqnarray}
with
\begin{eqnarray}
d_1(k) &=& 1+\cos k_x+\cos k_y,     \nonumber\\
d_2(k) &=& \sin k_x+\sin k_y,      \nonumber\\
d_3(k) &=& 2\lambda[\sin (k_x-k_y)-\sin k_x+\sin k_y]. \label{ghm}
\end{eqnarray}

We will work in the eigen-spin space and reduce the four-band problem to the two-band one. Here, we study the states only for spin-up electrons. The eigen-wave function $\psi(k)$ and energy $E(k)$ for a bulk state with momentum $k$ in valence band is given by
\begin{eqnarray}
y(k) &=& \begin{bmatrix}
	d_3(k)+E(k)\\
	d_1(k)+id_2(k)\\
\end{bmatrix}/N_k  \label{gwv1} \\
 E(k) &=& -\sqrt{d^2_1(k)+d^2_2(k)+d^2_3(k)}, \label{eng1}
\end{eqnarray}
where $N_k = \sqrt{2E(k)[E(k)+d_3(k)]}$ is the normalization constant.

For the present system, the longest hopping distance is $L = 1$ and the number of sites in the unit cell is $n = 2$. In the spin-up space, the matrix $h_j(k)$ defined by equation (\ref{bc1}) now is $h_1(k) = H(-1) = \tau_+  -i\lambda[1-\exp(ik_y)]$. For $k_y \ne 0$, the matrix $H(-1)$ is invertible. However, at $k_y = 0$, $H(-1) = \tau_+$ is not invertible, which results in the noninvertible of $A_c$ and thereby the uncertainty of the wave functions of scattering states.

For the edge states, we here consider only two cases of the parallel momentum at the TR invariant points $k_y =0$ and $\pi$ below. 
 
(i) Case of $k_y = 0$. As noted above, since the scattering states at $k_y = 0$ is not uniquely defined, the procedure for searching the edge states developed in section 3 is problematic. In this case, we investigate the edge states at $k_y = 0$ by taking the limit $k_y \to 0$. According to our analysis, first, we need to know the rank of matrix $A(\vec q)$ at small $k_y$ from the band structure of the bulk states. To first order of $k_y$, from equation (\ref{eng1}), the valence band is
\begin{eqnarray}
E(k) \approx -\sqrt{5+4\cos k_x+2\sin k_x\sin k_y}. \label{engy}
\end{eqnarray}
The momentum of the minimum is given by $k_x \approx k_y/2$. We thus have $m = 1$ for small $k_y$. In the limit $k_y \to 0$, the number $m = 1$ is apparently unchanged. Therefore, we can construct the matrix $A(\vec q)$ by firstly taking the limit $k_y \to 0$.

By noting that $h_1(q) = \tau_+$, the matrix $A(q)$ is obtained as
\begin{eqnarray}
A(q) &=& h_1(q)\begin{bmatrix}
	d_3(q)+E(q)\\
	d_1(q)+id_2(q)\\
\end{bmatrix}  \nonumber\\
&=& \begin{bmatrix} 
	d_1(q)+id_2(q)\\
	0\\
\end{bmatrix}.\nonumber
\end{eqnarray}
The matrix $M(q)$ is
\begin{eqnarray}
M(q) = d_1(q)+id_2(q) = z+2. \nonumber
\end{eqnarray}
Clearly, there is no zero of $\det M(q)$ in the unit circle and no edge state in the system.

By the way, here, we illustrate why one cannot assert the vanishing of the wave function at sites $1-L\le j\le 0$ from equation (\ref{bc}). In the present case, the boundary condition for the wave function with energy $E$ given by equation (\ref{bc}) reads
\begin{eqnarray}
H(-1)\psi(0;E) = 0. \label{eng2}
\end{eqnarray}
Here, $H(-1) = \tau_+$ at $k_y = 0$. Because $H(-1)$ is not invertible, a nonvanishing value of $\psi(0;E)$ is allowable.  

(ii) Case of $k_y = \pi$. Now, from equation (\ref{ghm}), we have   
\begin{eqnarray}
d_1(k) &=& \cos k_x,     \nonumber\\
d_2(k) &=& \sin k_x,      \nonumber\\
d_3(k) &=& -4\lambda \sin k_x. \nonumber 
\end{eqnarray}
The energy of valence band is given by  
\begin{eqnarray}
E(k)  = -\sqrt{1+16\lambda^2\sin^2 k_x}, \label{eng3}
\end{eqnarray}
which has the structure as shown in figure E2. 

\begin{figure}[t]
\centerline{\epsfig{file=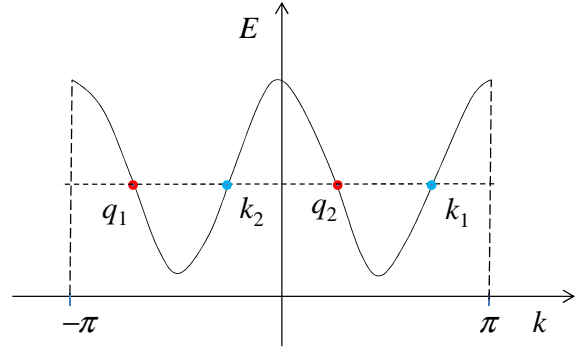,width=8.cm}}
\caption{Sketch of the energy band as function of momentum given by equation (\ref{eng3}). The wavenumbers of incoming and outgoing waves are denoted as red and blue dots, respectively. } 
\end{figure}

{\it Construction of matrix $A(\vec q)$.} The rank of the matrix $A(\vec q)$ is $m = 2$. The two wavenumbers for incoming waves are $q_1=q$ and $q_2=q+\pi$. According to the definition given by equation (\ref{wv}), the two wave functions $w(q_1)$ and $w(q_2)$ are
\begin{eqnarray}
w(q_1) = \begin{bmatrix}
	d_3(q)+E(q)\\
	z\\
\end{bmatrix},  
w(q_2) = \begin{bmatrix}
	-d_3(q)+E(q)\\
	-z\\
\end{bmatrix} \nonumber
\end{eqnarray}
where $z = \exp(iq)$. For $h(q)$, we have $h(q) = h_1(q) = H(-1) = \tau_+-i2\lambda\tau_z$. Since $h(q)$ now is a constant invertible matrix, we can drop it in constructing the matrix $A(\vec q)$. [The matrix $h(q)$ is equivalent to a row transformation on $Y(q)$ in the present case. Because $M(\vec q)$ is obtained from $A(\vec q)$ by row transformation, $h(q)$ can be dropped now. The zeros of $\det M(\vec q)$ are invariable under a row transformation.] The matrix $M(\vec q)$ is then obtained as
\begin{eqnarray}
M(\vec q) = \begin{bmatrix}
	d_3(q)+E(q)&-d_3(q)+E(q)\\
	z&-z\\
\end{bmatrix}. \nonumber
\end{eqnarray}
The determinant is given by
\begin{eqnarray}
\det M(\vec q) = -2zE(q).  \label{dtm}
\end{eqnarray}
The zeros of $\det M(\vec q)$ are 
\begin{eqnarray}
z_{s_1s_2} = s_1(1+s_2\sqrt{1+\Delta^2})/\Delta  \nonumber
\end{eqnarray}
with $\Delta = 4\lambda$ and $s_{1,2} = \pm$. The two zeros inside the unit circle are,
\begin{eqnarray}
z_{s_1-} = s_1(1-\sqrt{1+\Delta^2})/\Delta.  \label{zrsd}
\end{eqnarray}
Written in terms of $z$ and the zeros, $\det M(\vec q)$ is given by
\begin{eqnarray}
\det M(\vec q) = i\Delta \prod_{s_1,s_2}(z-z_{s_1s_2})^{1/2}.  \nonumber
\end{eqnarray}
Clearly, these zeros are irregular ones with $n_i = 2$. In the $z$ plane, the matrix $M(\vec q)$ is analytically defined on the two-sheet disk enclosed by the unit circle as shown in figure E3. The two red points are the irregular zeros. The two sheets of the disk are dissected between the two zeros. Along the dashed line between the red point on the right and (1,0), the first sheet is smoothly attached to the second sheet. 

\begin{figure}[t]
\centerline{\epsfig{file=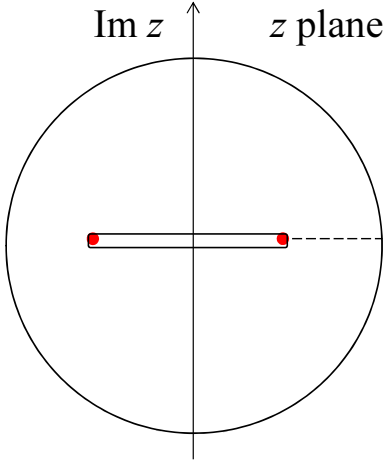,width=6.cm}}
\caption{Domain enclosed by the unit circle for analytically define $\det M(\vec q)$. The red points are the zeros of $\det M(\vec q)$ given by equation (\ref{zrsd}). Between the two red points, the domain is dissected. The red point on the right is the branch point with the dashed line as the cut-line.} 
\end{figure}

{\it Edge states.} The two zeros inside the unit circle are equivalent for determining the single SS. The vector $a$ is determined by
\begin{eqnarray}
M(\vec q)|_{z=z_{s_1-}}a = 0,  \nonumber
\end{eqnarray}
from which we get $a =(1,1)$. The energy is zero because of equation (\ref{dtm}). The wave function is obtained as
\begin{eqnarray}
\psi_b(j,0) = \frac{1}{N_0}\begin{pmatrix}
	i\\
	z_{+-}\\
\end{pmatrix}[1-(-1)^j]z^j_{+-},  \nonumber
\end{eqnarray}
where $N_0$ is a normalization constant. The wave function is nonzero only for the unit cells at the odd sites $j$ and is damping as $j$ going into the interior of the graphene lattice.

Lastly, we would like to illustrate the definition of the contour integral in equation (\ref{ns}) for the present problem. The contour $c$ can be defined as: starting from the point (1,0) on the first sheet of the disk, then running counterclockwise along the unit circle, finally getting to the point (1,0) on the second sheet.

\subsection{Edge states of electrons in graphene as semi-metal}

In the absence of the spin-orbit interactions, the conduction and valence bands touch at the Dirac points and graphene is a semi-metal without gap. The Hamiltonian reduces to
\begin{eqnarray}
H(k) = \begin{bmatrix}
	0&t_{-k}\\
	t_k&0\\
\end{bmatrix}, \nonumber
\end{eqnarray}
with 
\begin{eqnarray}
t_k = -[1+\exp(ik_x)+\exp(ik_y)]. \nonumber
\end{eqnarray}
Up to a normalization constant, the wave function of a bulk state in valence band is given by
\begin{eqnarray}
w(k) = \begin{bmatrix}
	t_{-k}\\
	E(k)\\
\end{bmatrix}, \nonumber
\end{eqnarray}
with the energy 
\begin{eqnarray}
E(k) = -[3+2\cos k_y+4\cos k_y/2 \cos(k_x-k_y/2)]^{1/2}. \nonumber\\
\label{engg}
\end{eqnarray}
In the present case, we have $m = 1$ for the rank of the matrix $A(\vec q)$. Because of $h(q) = H(-1) = \tau_+$, the matrix $A(q)$ is
\begin{eqnarray}
A(q) = h(q) w(q) = \begin{bmatrix}
	E(q)\\
	0\\
\end{bmatrix}, \nonumber
\end{eqnarray}
and thereby 
\begin{eqnarray}
M(q) = E(q).\nonumber
\end{eqnarray}
From equation (\ref{engg}), the maximum point of the energy is at $q = k_y/2+\pi$. According to our analysis, the zero point of $M(q)$ should be $q = k_y/2 +\pi+ iq_i$. Therefore, we have
\begin{eqnarray}
E^2(q)=3+2\cos k_y-4\cos k_y/2 \cosh(q_i) = 0,
\end{eqnarray}
from which we get
\begin{eqnarray}
q_i = -\log(2\cos k_y/2).
\end{eqnarray}
The condition that $q_i$ is positive requires $2\pi/3 < |k_y| <\pi$, in agreement with the existing work \cite {Fujita,Mong,Akhmerov}. The wave function is obtained as
\begin{eqnarray}
\psi_b(j,0) = \frac{1}{N_0}\begin{pmatrix}
	1\\
	0\\
\end{pmatrix}r^j\exp(ik_yj/2)  \nonumber
\end{eqnarray}
where $r = -2\cos k_y/2$ and $N_0$ is a normalization constant.

\subsection{Edge states in a $d$-wave superconductor}

We consider a two-dimensional $d$-wave superconductor with a (11) edge as shown in figure E4. The effective Hamiltonian for electrons is given by 
\begin{eqnarray}
H = \sum_{\langle ij\rangle k_y}C^{\dagger}_i(k_y)H_{ij}(k_y)C_j(k_y) \label{hms}
\end{eqnarray}
where $k_y$ is the momentum parallel to the edge, $C^{\dagger}_i (k_y)= (c^{\dagger}_{i,k_y,\uparrow},c_{i,-k_y,\downarrow})$. Here, $h_{ij}(k_y)$'s are given by 
\begin{eqnarray}
H_{jj}(k_y) &\equiv& H(0) \nonumber\\
&=& \begin{bmatrix}
	-\mu&0\\
	0&\mu\\
\end{bmatrix}\nonumber\\
H_{j,j+1}(k_y) &\equiv& H(1) \nonumber\\
&=& \begin{bmatrix}
	-(1+e^{-ik_y})&\Delta(1-e^{-ik_y})\\
	\Delta(1-e^{-ik_y})&(1+e^{-ik_y})\\
\end{bmatrix}\nonumber\\
H_{j,j-1}(k_y) &\equiv& H(-1)\nonumber\\
&=&\begin{bmatrix}
	-(1+e^{ik_y})&\Delta(1-e^{ik_y})\\
	\Delta(1-e^{ik_y})&(1+e^{ik_y})\\
\end{bmatrix}\nonumber
\end{eqnarray}
where $\mu < 0$ is the chemical potential, and $\Delta$ is the pairing order parameter. This is a tight-binding model with $L = 1$.

\begin{figure}[t]
\centerline{\epsfig{file=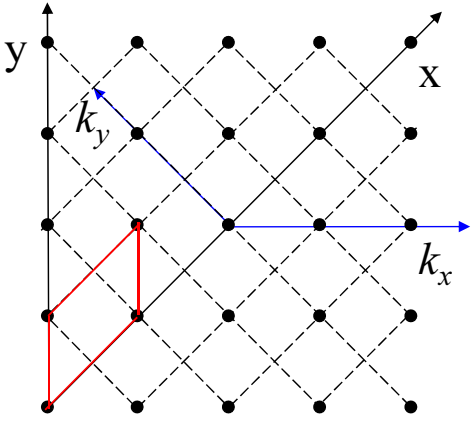,width=7.cm}}
\caption{Square lattice with a (11) edge. The red quadrilateral is a unit cell. The axes of real $(x,y)$ and reciprocal $(k_x,k_y)$ spaces are also shown.}
\end{figure}

The Hamiltonian for the bulk states in momentum space is given by
\begin{eqnarray}
H(k) =\begin{pmatrix}
	\xi_q&\Delta_q\\
	\Delta_q&-\xi_q\\
\end{pmatrix}\nonumber
\end{eqnarray}
with $\xi_q=-4\cos q\cos k_y/2-\mu, \Delta_q = -4\Delta\sin q\sin k_y/2$ and $q= k_x-k_y/2$. The energy and eigen-wave function of the excited quasiparticles are  
\begin{eqnarray}
E(q) &=&\sqrt{\xi^2_q+\Delta^2_q},   \label{edwsc}\\
\psi(q)&=&\frac{1}{N_q} \begin{bmatrix}
	\xi_q+E(q)\\
	\Delta_q\\
\end{bmatrix}. \nonumber
\end{eqnarray}
For brevity, we have suppressed the argument $k_y$. A sketch of the energy as function of shifted momentum $q$ is shown in figure E5. The minimum point $q_0$ satisfies
\begin{eqnarray}
\cos q_0 = -\frac{c_1\mu}{c_1^2-c_2^2} \label{cdq0}
\end{eqnarray}
with $c_1=4\cos k_y/2$, and $c_2= 4\Delta\sin k_y/2$. From the band structure, we have $m = 2$ for $k_y\ne 0$. The wavenumbers of two incoming waves $q_1$ and $q_2$ are related by
\begin{eqnarray}
\cos q_1 +\cos q_2= 2\cos q_0. \nonumber
\end{eqnarray}

\begin{figure}[t]
\centerline{\epsfig{file=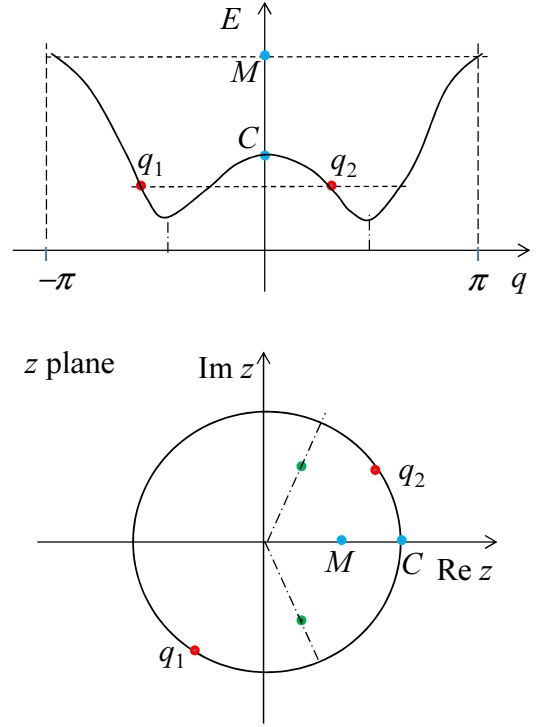,width=8.cm}}
\caption{Top panel: Sketch of the positive energy band given by equation (\ref{edwsc}) as function of momentum. The wavenumbers of incoming waves are denoted as red points. Lower panel: $z$ plane for the mapping $z = \exp(iq)$. The green points on the dash-dot lines in the unit circle are the zeros of $\det M(\vec q)$.} 
\end{figure} 

{\it Construction of matrix $A(\vec q)$.} For studying the edge states, we start to construct matrix $A(\vec q)$. Here, again $h(q) = H(-1)$ is independent on $q$ and is invertible. Therefore, we disregard $h(q)$ for constructing $A(\vec q)$. In this case, the matrix $M(\vec q)$ is obtained as
\begin{eqnarray}
M(\vec q) = \begin{bmatrix}
	\xi_q+E(q)&\xi_{q_2}+E(q)\\
	\Delta_q&\Delta_{q_2}\\
\end{bmatrix}, \nonumber
\end{eqnarray}
where $q = q_1$. We have
\begin{eqnarray}
\det M(\vec q) =\Delta_{q_2}[\xi_q+E(q)]-\Delta_{q}[\xi_{q_2}+E(q)]. \nonumber
\end{eqnarray}
According to the analysis described in section 3, the zeros of $\det M(\vec q)$ are placed along the dash-dot lines (for energy below the minima of the energy band) in lower panel of figure E5. Because these points correspond to the same energy, they are complex conjugates of each other. Therefore, $\Delta_{q_2} = - \Delta^{\ast}_{q}$, and $\xi_{q_2} = \xi^{\ast}_{q}$. For the zeros of $\det M(\vec q)$, we then arrive at
\begin{eqnarray}
\det M(\vec q) =-2{\rm Re}\{\Delta_{q}[\xi^{\ast}_{q}+E(q)]\} = 0, \nonumber
\end{eqnarray}
which can be satisfied with
\begin{eqnarray}
\xi_{q} =-i\lambda\Delta_{q}, ~~~\lambda = \pm 1.\label{sfdw}
\end{eqnarray}
From equation (\ref{sfdw}), we obtain
\begin{eqnarray}
E(q) &=& 0,  \nonumber\\
z_0 &=&\exp(iq)= r\exp(iq_0) ,\nonumber\\
r &=& \sqrt{\frac{c_1-|c_2|}{c_1+|c_2|}} ,\label{zrdw}
\end{eqnarray}
and $\lambda = {\rm sgn} (c_2)$. 

{\it Edge state.} From equation (\ref{sfdw}), the matrix $M(q)$ becomes
\begin{eqnarray}
M(\vec q) = \begin{bmatrix}
	-i\lambda\Delta_q&i\lambda\Delta^{\ast}_q\\
	\Delta_q&-\Delta^{\ast}_{q}\\
\end{bmatrix}. \nonumber
\end{eqnarray}
The vector $a$ is determined by 
\begin{eqnarray}
M(\vec q)a = 0, \nonumber
\end{eqnarray}
which gives rise to $a = (1,\Delta_q/\Delta^{\ast}_{q})$. For the wave function, we get
\begin{eqnarray}
\psi_b(j,0) = \frac{1}{N_0}\begin{bmatrix}
	1\\
	i{\rm sgn} (c_2)\\
\end{bmatrix}r^j\sin(q_0j), \label{wvfdw}
\end{eqnarray}
where $N_0$ is a normalization constant. This is the known result \cite{Yan1}.

\end{document}